\documentclass[5p,onecolumn]{elsarticle}

\usepackage[cmex10]{amsmath}
\usepackage{rotating}
\usepackage{ifpdf}
\usepackage{algorithmic}
\usepackage{graphicx}
\usepackage{multirow}
\usepackage[TABBOTCAP]{subfigure}
\usepackage{caption}
\usepackage{subfigure}
\usepackage{amsfonts}
\usepackage{tabularx}
\usepackage{float}
\usepackage{balance}
\newfloat{Code}{H}{myc}
\usepackage{latexsym}
\usepackage{amssymb}
\usepackage{lscape}
\usepackage{colortbl}
\usepackage{url}
\usepackage{enumitem}
\usepackage{color}
\usepackage[framemethod=tikz]{mdframed}
\usepackage{adjustbox}
\definecolor{sgray}{gray}{0.5}
-lmr10

\newcommand{\ie}{\textit{i.e.,}\space}
\newcommand{\eg}{\textit{e.g.,}\space}

\newcommand{\etal}{\textit{et al.}\space}

{

\usepackage{listings}
\lstset{language=Java}
\definecolor{mygreen}{rgb}{0,0.6,0}
\definecolor{mygray}{rgb}{0.5,0.5,0.5}
\definecolor{mymauve}{rgb}{0.58,0,0.82}
\lstset{ %
  backgroundcolor=\color{white},   
  basicstyle=\footnotesize,        
  breakatwhitespace=false,         
  breaklines=true,                 
  captionpos=b,                    
  commentstyle=\color{mygreen},    
  deletekeywords={...},            
  escapeinside={\%*}{*)},          
  extendedchars=true,              
  frame=single,                    
  keepspaces=true,                 
  keywordstyle=\color{blue},       
  language=Octave,                 
  morekeywords={*,...},            
  numbers=left,                    
  numbersep=5pt,                   
  numberstyle=\tiny\color{mygray}, 
  rulecolor=\color{black},         
  showspaces=false,                
  showstringspaces=false,          
  showtabs=false,                  
  stepnumber=2,                    
  stringstyle=\color{mymauve},     
  tabsize=2,                       
  title=\lstname                   
}

\definecolor{gray50}{gray}{.5}
\definecolor{gray40}{gray}{.6}
\definecolor{gray30}{gray}{.7}
\definecolor{gray20}{gray}{.8}
\definecolor{gray10}{gray}{.9}
\definecolor{gray05}{gray}{.95}

\newlength\Linewidth
\def\findlength{\setlength\Linewidth\linewidth
	\addtolength\Linewidth{-4\fboxrule}
	\addtolength\Linewidth{-3\fboxsep}
}

\newenvironment{examplebox}{\par\begingroup
	\setlength{\fboxsep}{5pt}\findlength
	\setbox0=\vbox\bgroup\noindent
	\hsize=0.95\linewidth
	\begin{minipage}{0.95\linewidth}\normalsize}
	{\end{minipage}\egroup
	\textcolor{gray20}{\fboxsep1.5pt\fbox
		{\fboxsep5pt\colorbox{gray05}{\normalcolor\box0}}}
	\endgroup\par\noindent
	\normalcolor\ignorespacesafterend}

\newenvironment{resultbox}{\par\begingroup
	\setlength{\fboxsep}{5pt}\findlength
	\setbox0=\vbox\bgroup\noindent
	\hsize=0.95\linewidth
	\begin{minipage}{0.95\linewidth}\normalsize}
	{\end{minipage}\egroup
	\textcolor{gray20}{\fboxsep1.5pt\fbox
		{\fboxsep5pt\colorbox{white}{\normalcolor\box0}}}
	\endgroup\par\noindent
	\normalcolor\ignorespacesafterend}

\newenvironment{commitbox}{\par\begingroup
	\setlength{\fboxsep}{5pt}\findlength
	\setbox0=\vbox\bgroup\noindent
	\hsize=0.95\linewidth
	\begin{minipage}{0.95\linewidth}\normalsize}
	{\end{minipage}\egroup
	\textcolor{gray20}{\fboxsep1.5pt\fbox
		{\fboxsep5pt\colorbox{white}{\normalcolor\box0}}}
	\endgroup\par\noindent
	\normalcolor\ignorespacesafterend}

\usepackage[framemethod=tikz]{mdframed}
\newmdenv[innerlinewidth=0.5pt, roundcorner=4pt,innerleftmargin=6pt,
innerrightmargin=6pt,innertopmargin=6pt,innerbottommargin=6pt]{mybox}

\input{macro}
\newcommand\revised[1]{\textcolor{black}{#1}}

\sloppy
\begin{document}

\begin{frontmatter}

\markboth{Catolino et al.}{Not All Bugs Are the Same: Understanding, Characterizing, and Classifying the Root Cause of Bugs}

\title{\revised{Not All Bugs Are the Same:\\Understanding, Characterizing, and Classifying the Root Cause of Bugs}}

\author{Gemma Catolino$^{1}$, Fabio Palomba$^{2}$, Andy Zaidman$^{3}$, Filomena Ferrucci$^{1}$}
\address{$^{1}$University of Salerno, Italy --- $^{2}$University of Zurich, Switzerland --- $^{3}$Delft University of Technology, The Netherlands\\
gcatolino@unisa.it, palomba@ifi.uzh.ch, a.e.zaidman@tudelft.nl, fferrucci@unisa.it}

	\begin{abstract}
	Modern version control systems such as \textsc{Git} or \textsc{SVN} include bug tracking mechanisms, through which developers can highlight the presence of 
	bugs through bug reports, \ie textual descriptions reporting the problem and what are the steps that led to a failure. In past and recent years, the research community 
	deeply investigated methods for easing bug triage, that is, the process of assigning the fixing of a reported bug to the most qualified developer. Nevertheless, only a 
	few studies have reported on how to support developers in the process of understanding the type of a reported bug, which is the first and most time-consuming step 
	to perform before assigning a bug-fix operation. In this paper, we target this problem in two ways: first, we analyze 1,280 bug reports of 119 popular projects belonging 
	to three ecosystems such as \textsc{Mozilla}, \textsc{Apache}, and \textsc{Eclipse}, with the aim of building a taxonomy of the root causes of reported bugs; then, we 
	devise and evaluate an automated classification model able to classify reported bugs according to the defined taxonomy. As a result, we found nine main common 
	root causes of bugs over the considered systems. Moreover, our model achieves high F-Measure and AUC-ROC (64\% and 74\% on overall, respectively).	
\end{abstract}

	
\end{frontmatter}



\section{Introduction}
\label{sec:intro}

The year 2017 has been earmarked as \emph{The Year That Software Bugs Ate The World}.\footnote{\url{https://tinyurl.com/y8n4kxgw}, last visited April 17th, 2018.} It serves as an apt reminder that software engineers are but human, and have their fallacies when it comes to producing bug-free software. With modern software systems growing in size and complexity, and developers having to work under frequent deadlines, the introduction of bugs does not really come as a surprise.

Users of such faulty software systems have the ability to report back software failures, either through dedicated issue tracking systems or through version control platforms like GitHub. In order to do so, a user files a so-called \emph{bug report}, which contains a textual description of the steps to perform in order to reproduce a certain failure~\cite{breu2010information,hooimeijer2007modeling}. 

	
%
%
		Once a failure is known and reported, the bug localization and fixing process starts~\cite{DBLP:books/daglib/0039904,DBLP:conf/icse/BellerSSZ18}. Developers are requested to (i) analyze the bug report, (ii) identify the root cause of the bug, \eg if it is a security- or a performance-related one, and (iii) assign its verification and resolution to the most qualified developer \cite{zhang2013hybrid}. The research community proposed methodologies and tools to identify who should fix a certain bug \cite{anvik2006automating,anvik2006should,anvik2011reducing,jeong2009improving,murphy2004automatic,xuan2015towards,xuan2017automatic}, thus supporting developers once they have diagnosed the root cause of the problem they have to deal with. 
		
		However, there is still a lack of approaches able to support developers while analyzing a bug report in the first instance.
		As a matter of fact, understanding the root cause of a bug represents the first and most time-consuming step to perform in the process of bug triage \cite{akila2015effective}, since it requires an in-depth analysis of the characteristics of a newly reported bug report. Unfortunately, such a step is usually performed manually before the assignment of a developer to the bug fix operation \cite{breu2010information}. Perhaps more importantly, most of the research approaches aimed at supporting the bug triage process treat all bugs in the same manner, without considering their root cause \cite{zaman2011security}. 
		
		\begin{center}
			\begin{resultbox}
				\emph{We argue that the definition of approaches able to support developers in the process of understanding the root cause of bugs can be beneficial to properly identify the developer who should be assigned to its debugging, speeding-up the bug analysis and resolution process.}
			\end{resultbox}
		\end{center}
	
		\begin{figure*}
			\centering
			\includegraphics[width=1\linewidth]{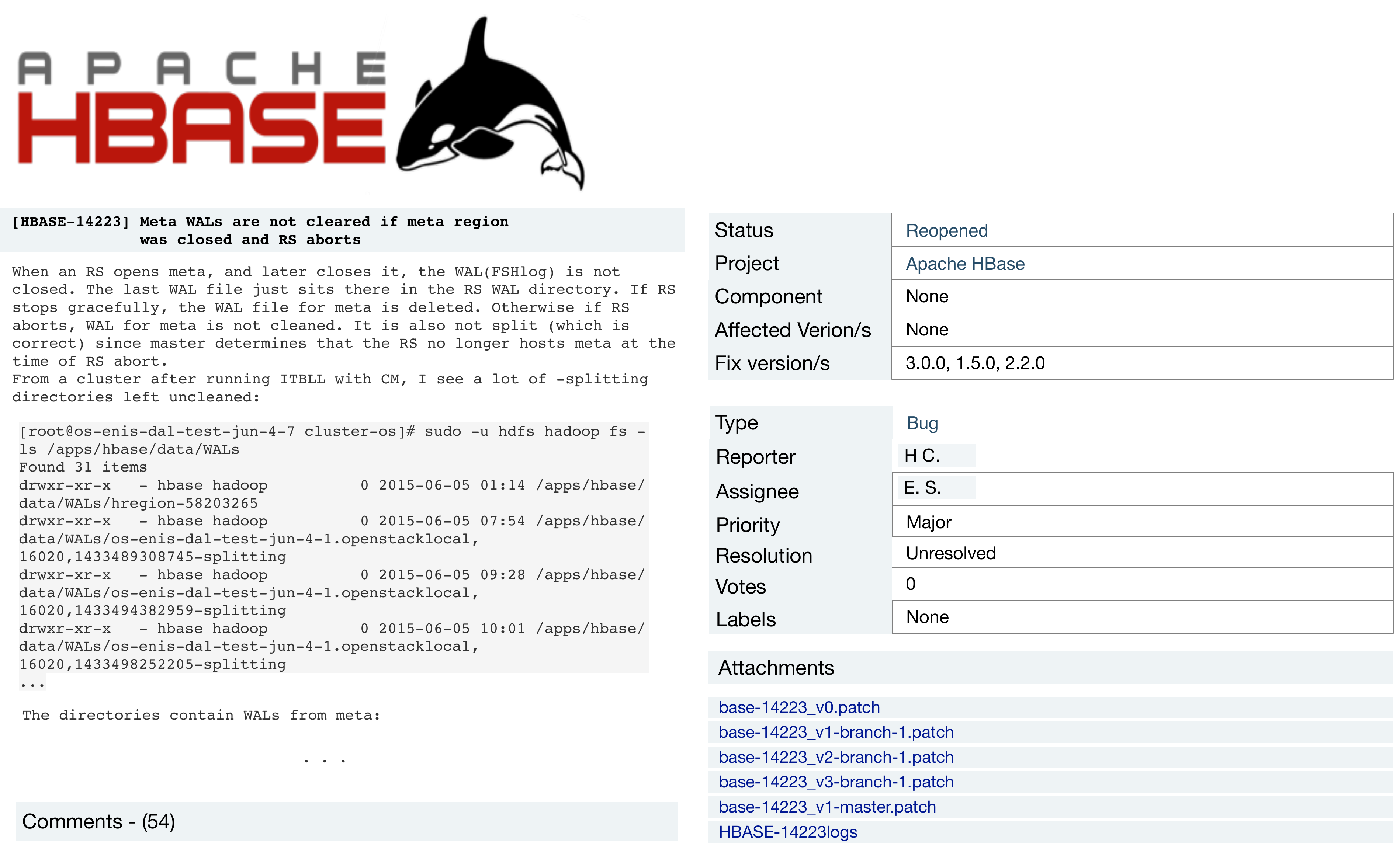}
			\caption{\revised{Bug reported and reopened in \textsc{Apache HBase}.}}
			\label{fig:bugReportHBaseExample}
		\end{figure*}
		
		\subsection{Motivating example} 
		\revised{To support our statement, let us consider a real bug report from \textsc{Apache HBase},\footnote{https://hbase.apache.org} one of the projects taken into account in our study. This project implements a scalable distributed big data store able to host large relational tables atop clusters of commodity hardware. On August 15th, 2015 the bug report shown in Figure \ref{fig:bugReportHBaseExample} was created.\footnote{The full version of the bug report is available here: https://goo.gl/rS8iQU.}}
		
		\revised{The developer who opened the report (\ie H. C.) encountered an issue related to a network-related problem. Specifically, due to the wrong management of the so-called World Atlas of Language Structures (WALS), a large set of structural (phonological, grammatical, lexical) properties of languages gathered from descriptive materials (such as reference grammars). Specifically, when a Region Server (RS) aborts its operations, the directory containing the WAL data is not cleaned, causing possible data incoherence or inconsistency issues. Looking at the change history information of the system, the class \texttt{HRegionServer}---the file containing the reported bug---has been mostly modified by developer G. C., who was indeed assigned to the resolution of this bug report on May 30th, 2017. Such an assignment is in line with the recommendations provided by existing bug triaging approaches \cite{shokripour2013so,tian2016learning}, that would suggest G. C. as an assignee since he has a long experience with this class.
		However, \emph{not all bugs are the same}: a more careful analysis of the types of changes applied by G. C. reveals that they were mainly focused on the configuration of the server rather than on the communication with the client. As a result, the bug was marked as \textsf{`resolved'} on September 17th, 2017: however, the bug was not actually fixed and was reopened on October 6th, 2018. This indicates that the experience of a developer on a certain class---taken as relevant factor within existing bug triaging approaches \cite{shokripour2013so,tian2016learning}---might not be enough for recommending the most qualified developer to fix a bug. In other words, we conjecture that understanding the root cause of a newly reported bug might be beneficial for bug triage. At the same time, it might reduce the phenomenon of bug tossing \cite{jeong2009improving}---which arises when developers re-assign previously assigned bugs to others, as in the example above---by allowing a more correct bug assignment.} 
			
		\subsection{Our work and contributions} 
		In this paper, we aim to perform the first step towards the (i) empirical understanding of the possible root causes behind bugs and (ii) automated support for their classification. To that end, we first propose a novel taxonomy of bug root causes, that is built on the basis of an iterative content analysis conducted on 1,280 bug reports of 119 software projects belonging to three large ecosystems such as \textsc{Mozilla}, \textsc{Apache}, and \textsc{Eclipse}. To the best of our knowledge,  this is the first work that proposes a general taxonomy collecting the main root causes of software bugs. This also enables the construction of a dataset of labeled bug reports that can be further exploited. 
		
		In the second place, we build an automated approach to classify bug reports according to their root cause; in other words, we built our classification model training our classifier using the textual content of the bug report to predict its root cause.
		
		We empirically evaluate the classification model by running it against the dataset coming as output of the taxonomy building phase, measuring its performance adopting a 100 times 10-fold cross validation methodology in terms of F-Measure, AUC-ROC, and Matthew's Correlation Coefficient (MCC). 
		
		The results of the study highlight nine different root causes behind the bugs reported in bug reports, that span across a broad set of issues (\eg GUI-related vs Configuration bugs) and are widespread over the considered ecosystems. In addition, the classification model shows promising results, as it is able to classify the root causes of bugs with an F-Measure score of 64\%.
		
		To sum up, the contributions made by this paper are:
		
		\begin{enumerate}
			
			\item A \emph{taxonomy} reporting the common root causes of bugs raised through bug reports, and that has been manually built considering a large corpus of existing bug reports;
			
			\item An in-depth analysis of the characterization of the different bug types discovered. In particular, we took into account three different perspectives such as frequency, relevant topics discussed, and the time needed to fix each bug type.
			
			\item A novel \emph{root cause classification model} to automatically classify reported bugs according to the defined taxonomy.
			
			\item A large \emph{dataset} and a replication package \cite{appendix} that can be used by the research community to further study the characteristics of bug reports and bugs they refer to.
			
		\end{enumerate}
	
	The remainder of the paper is as follows. Section \ref{sec:related} overviews the related literature on bug report analysis and classification. Section \ref{sec:study} describes the research methodology adopted to conduct our study, while in Section \ref{sec:results} we report the achieved results. In Section \ref{sec:discussion} we deeper discuss our findings and the implications of our work. Section \ref{sec:threats} examines the threats to the validity of the study and the way we mitigated them. Finally, Section~\ref{sec:conclusions} concludes the paper and provides insights into our future research agenda.

\section{Background and Related Work}
\label{sec:related}
Our work revolves around the problem of classifying bugs according to their root cause with the aim of supporting and possibly speeding-up bug triaging activities. Thus, we focus this section on the discussion of the literature related to bug classification studies and approaches. A comprehensive overview of the research conducted in the context of bug triaging is presented by Zhang \etal \cite{zhang2016literature}.
		
		\subsection{Bug classification schemas}
			The earliest and most popular bug classification taxonomy was proposed by \textsc{IBM} \cite{chillarege1992orthogonal}, which introduced the so-called \emph{Orthogonal Defect Classification} (ODC). This taxonomy includes 13 categories that allow developers to separate bugs depending on their impact on the customer. Thus, the focus is on the \emph{effect} produced by bugs rather than on their \emph{root cause}.
			
			Another popular bug characterization schema was developed by Hewlett-Packard \cite{freimut2005industrial}. In this case, bugs are characterized by three attributes: (i) ``origin'', that is the activity in which the defect was introduced (\eg during requirement specification or design); (ii) ``mode'', which describes the scenarios leading to a bug; and (iii) ``type'', that describes more in-depth the origin of a bug, by specifying if it is hardware- or software-related. It is important to note that the attribute ``type'' of this classification schema is not intended to be used for the specification of the root cause of a bug (\eg a performance issue), but rather it provides more context on the location of a bug. 
			
			More recent works defined \emph{ad-hoc} taxonomies (i) for specific application types or (ii) aiming at characterizing particular root causes of bugs. 
			As for the former category, Chan \etal~\cite{chan2007fault} proposed a taxonomy that captures the possible failures that arise in Web service composition systems, while Bruning \etal~\cite{bruning2007fault} provided a corresponding fault taxonomy for service-oriented architectures according to the process of service invocation. Similarly, Ostrand \etal\cite{ostrand1984collecting} conducted an industrial study involving an interactive special-purpose editor system, where a group of developers were asked to categorize 173 bugs based on the error they referred to: as a final result, a taxonomy was built. Lal and Sureka \cite{lal2012comparison} analyzed commonalities and differences of seven different types of bug reports within Google; they provided guidelines to categorize their bug reports. Moreover, recent studies \cite{leszak2002classification,buglione2006introducing} showed how developers manually classify defects into the ODC categories based on the reported descriptions using, for example, root cause defect analysis. 
						
			As for the second category (related to the characterization of particular root causes of bugs), Aslam \etal \cite{aslam1996use} defined a classification of security faults in the Unix operating system. More recently, \revised{Zhang \etal \cite{zhang2018empirical} analyzed the symptoms and root causes of  175 TensorFlow coding bugs from GitHub issues and StackOverflow questions. As a result, they proposed a number of challenges for their detection and localization.} Tan \etal \cite{tan2014bug} proposed work closest to ours: they started from the conjecture that three root causes, \ie semantic, security, and concurrency issues, are at the basis of most relevant bugs in a software system. Thus, they investigated the distribution of these bug root causes in projects such as \textsc{Apache}, \textsc{Mozilla}, and \textsc{Linux}. Finally, they performed a fine-grained analysis on the impact and evolution of such bugs on the considered systems; they proposed a machine learning approach using all the information of a bug report to automatically classify these semantic, security, and concurrency bugs and having an average F-Measure of $\approx70\%$. As opposed to the work by Tan \etal \cite{tan2014bug}, we start our investigation without any initial conjecture: as such, we aim at providing a wider overview of the root causes of bugs and their diffusion on a much larger set of systems (119 vs 3); furthermore, we aim at producing a high-level bug taxonomy that is \emph{independent} from the specific type of systems, thus being generically usable. Finally, the presented root cause classification model is able to  automatically classify all the identified root causes of bugs, thus providing a wider support for developers.
			
		\subsection{Bug classification techniques}
			Besides classification schemas, a number of previous works devised automated approaches for classifying bug reports. Antoniol \etal \cite{antoniol2008bug} defined a machine learning model to discriminate between bugs and new feature requests in bug reports, reporting a precision of 77\% and a recall of 82\%. In our case, we only consider bug reports actually reporting issues of the considered applications, since our goal is to classify bugs.
			Hern{\'a}ndez-Gonz{\'a}lez \etal \cite{hernandez2018learning} proposed an approach for classifying the impact of bugs according to the ODC taxonomy \cite{chillarege1992orthogonal}: the empirical study conducted on two systems, \ie \emph{Compendium} and \emph{Mozilla}, showed good results.  
			\revised{At same time, Huang \etal \cite{huang2015autoodc}, based on the ODC classification, proposed \textsc{AutoODC}, an approach for automating ODC classification by casting it as a supervised text classification problem and integrating experts' ODC experience and domain knowledge; they built two models trained with two different classifiers such as Naive Bayes and Support Vector Machine on a larger defect list extracted from FileZilla. They reported promising results. With respect to this work, our paper aims at providing a more comprehensive characterization of the root-causes of software bugs, as well as providing an automated solution for tagging them.}
	
			Thung \etal \cite{thung2012automatic} proposed a classification-based approach that can automatically classify defects into three super-categories that are comprised of ODC categories: \emph{control and data flow}, \emph{structural}, and \emph{non-functional}. \revised{In a follow-up work \cite{thung2015active}, they extended the defect categorization. In particular, they combined clustering, active learning and semi-supervised learning algorithms to automatically categorize defects; they firstly picked an initial sample, extract the examples that are more informative for training the classification model, and incrementally refining the trained model. They evaluated their approach on 500 defects collected from JIRA repositories of three software systems. Xia \etal \cite{xia2014automatic} applied a text mining technique in order to categorize defects into fault trigger categories by analyzing the natural-language description of bug reports, evaluating their solution on 4 datasets, \eg Linux, Mysql, for a total of 809 bug reports.} Nagwani \etal \cite{nagwani2013generating} proposed an approach for generating the taxonomic terms for software bug classification using LDA, while Zhou \etal \cite{zhou2016combining} combined text mining on the defect descriptions with structured data (\eg priority and severity) to identify corrective bugs. Furthermore, text-categorization based machine learning techniques have been applied for bug triaging activities \cite{murphy2004automatic,javed2012automated} with the aim of assigning bugs to the right developers. \revised{With respect to the works mentioned above, our paper reinforces the idea of using natural language processing to automatically identify the root-causes of bugs; nevertheless, we provide a more extensive empirical analysis of the types of bugs occurring in modern software systems, as well as their categorization according to different perspectives such as frequency, relevant topics, and time required to be fixed.}
			
			On the basis of these works, in the context of our research we noticed that there is a lack of studies that try to provide automatic support for the labeling of bugs according to their root cause: for this reason, our work focuses on this aspect and tries to exploit a manually built taxonomy of bug root causes to accomplish this goal.


\section{Research Methodology} 
\label{sec:study}
In this section, we report the empirical study definition and design that we follow in order to create a bug root cause taxonomy and provide a root cause classification model.
	
	\subsection{Research Questions}
	\revised{The \emph{goal} of the study is threefold: (i) \emph{understanding} which types of bugs affect modern software systems, (ii) \revised{characterizing} them to better describe their nature, and (iii) \emph{classifying} bug reports according to their root cause. The \emph{purpose} is that of easing the maintenance activity related to bug triaging, thus improving the allocation of resources, \eg assigning a bug to the developer that is more qualified to fix a certain type of issue. 
	The \emph{quality focus} is on the comprehensiveness of the bug root cause taxonomy as well as on the accuracy of the model in classifying the root causes of bugs. 
	The \emph{perspective} is that of both researchers and practitioners: the former are interested in a deeper understanding of the root causes of bugs occurring in software systems, while the latter in evaluating the applicability of root cause prediction models in practice.
	The specific research questions formulated in this study are the following:} \smallskip
		
		\begin{itemize}
			
			\item \textbf{RQ$_1$} \emph{To what extent can bug root causes be categorized through the information contained in bug reports?}\medskip
			
			\item \revised{\textbf{RQ$_2$} \emph{What are the characteristics, in terms of frequency, topics, and bug fixing time, of different bug categories?}}\medskip
			
			\item \textbf{RQ$_3$} \emph{How effective is our classification model in classifying bugs according to their root cause exploiting bug report information?}\smallskip
			
		\end{itemize}

		In \textbf{RQ$_1$} our goal is to categorize the bug root causes through the analysis of bug reports that are reported in bug tracking platforms.
		\revised{Secondly, in \textbf{RQ$_2$} we analyze (i) frequency, (ii) relevant topics, and (iii) bug fixing time of each category with the aim of characterizing them along these three perspectives.} Finally, in \textbf{RQ$_3$} we investigate how effectively the categories of bug root causes can be automatically classified starting from bug reports via standard machine learning techniques, so that developers and project managers can be automatically supported during bug triaging.
		In the following subsections, we detail the design choices that allow us to answer our research questions. 

		\subsection{Context Selection}
			In order to answer our research questions, we first needed to collect a large number of bug reports from existing software projects. 
			To this aim, we took into account bug reports of three software ecosystems such as \textsc{Mozilla},\footnote{\url{https://bugzilla.mozilla.org}} \textsc{Apache},\footnote{\url{https://bz.apache.org/bugzilla/}} and \textsc{Eclipse}.\footnote{\url{https://bugs.eclipse.org/bugs/}} 
			The selection of these systems was driven by the results achieved in previous studies \cite{bettenburg2007quality,sun2010discriminative,zimmermann2010makes}, which reported the high-quality of their bug reports in terms of completeness and understandability.
			We randomly sampled 1,280 bug reports that were \textsf{`fixed'} and \textsf{`closed'}: as also done in previous work \cite{tan2014bug}, we included them because they have all the information required for understanding the root cause of bugs (\eg developers' comments or attachments).
			It is important to note that we checked and excluded from the random sampling the so-called \emph{misclassified} bug reports, \ie those that do not contain actual bugs~\cite{antoniol2008bug,herzig2013s}, by exploiting the guidelines provided by Herzig \etal \cite{herzig2013s}. In the end, our dataset is composed of bug reports from 119 different projects of the considered ecosystems.
			
			Table \ref{tab:systems} contains for each ecosystem the (i) number of projects we considered, and (ii) number of bug reports taken into account. The final dataset is available in our online appendix \cite{appendix}.
			
				\begin{table}[h]
				\caption{Characteristics of Ecosystems in Our Dataset}
				\label{tab:systems}
				\centering
				\resizebox{0.75\linewidth}{!}{
					\begin{tabular}{lcc}\hline
						\cellcolor[gray]{0.85}{Ecosystem} & \cellcolor[gray]{0.85}{Project} & \cellcolor[gray]{0.85}{Bug Reports}\\ \hline
						Apache & 60 & 406\\
						\cellcolor[gray]{0.9}{Eclipse} & \cellcolor[gray]{0.9}{39} & \cellcolor[gray]{0.9}{444} \\
						Mozilla & 20 &430\\\hline
						\rowcolor[gray]{.0} \textbf{\textcolor{white}{Overall}} & \textbf{\textcolor{white}{119}} & \textbf{\textcolor{white}{1,280}}\\
					\end{tabular}
				}
			\end{table}

		\subsection{\textbf{RQ$_1$}. Toward a Taxonomy of Bug Root Causes}\label{sec:categorization}
			To answer our first research question, we conducted three iterative \emph{content analysis sessions} \cite{Lidw2010a} involving two software engineering researchers, both authors of this paper, ($1$ graduate student and $1$ research associate) with at least seven years of programming experience. From now on, we refer to them as \emph{inspectors}. Broadly speaking, this methodology consisted of reading each bug report (both title and summary, which reports its detailed description), with the aim of assigning a label describing the root cause that the reported problem refers to. It is important to note that in cases where the bug report information was not enough to properly understand the root cause of the bug, we also considered patches, attachments, and source code of the involved classes, so that we can better contextualize the cause of the bug by inspecting the modifications applied to fix it. 
			The final goal was to build a \emph{taxonomy} representing the bug root causes that occur during both software development and maintenance. In the following, we describe the methodology followed during the three iterative sessions, as well as how we validate the resulting taxonomy.\smallskip
	
			\subsubsection{Taxonomy Building} Starting from the set of 1,280 bug reports composing our dataset, overall, each inspector \emph{independently} analyzed 640 bug reports.
	
			\begin{description}[leftmargin=0.3cm]
				\item [Iteration 1:] The inspectors analyzed an initial set of 100 bug reports. Then, they opened a discussion on the labels assigned to the root causes identified so far and tried to reach consensus on the names and meaning of the assigned categories. The output of this step was a draft taxonomy that contains some obvious categories (\eg security bugs), while others remain undecided.
				
				\item [Iteration 2:] The inspectors firstly re-categorized the 100 initial bug reports according to the decisions taken during the discussion, then used the draft taxonomy as basis for categorizing another set of 200. This phase was for both assessing the validity of the categories coming from the first step (by confirming some of them and redefining others) and for discovering new categories. After this step was completed, the inspectors opened a new discussion aimed at refining the draft taxonomy, merging overlapping root cause categories or characterizing better the existing ones. A second version of the taxonomy was produced.
				
				\item [Iteration 3:] The inspectors re-categorized the 300 bug reports previously analyzed. Afterward, they completed the final draft of the taxonomy verifying that each kind of bug root cause encountered in the final 339 bug reports was covered by the taxonomy.
			\end{description}
	
		Following this iterative process, we defined a taxonomy composed of 9 categories. It is important to note that at each step we computed the inter-rater agreement using the Krippendorff's alpha Kr$_\alpha$ \cite{bauer2007content}. During the sessions, the agreement measure ranged from 0.65, over 0.76, to 0.96 for the three iterative sessions, respectively. Thus, we can claim that the agreement increased over time and reached a considerably higher value than the 0.80 standard reference score usually considered for Kr$_\alpha$~\cite{antoine2014weighted}.
		
		\smallskip
		\textbf{Taxonomy Validation.} While the iterative content analysis makes us confident about the comprehensiveness of the proposed taxonomy, we also evaluated it in an alternative way: specifically, we involved 5 industrial developers having more than 10 years of programming experience. They were all contacted via e-mail by one of the authors of this paper, who selected them from her personal contacts.
		
		We provided them with an \textsc{Excel} file that contained a list of 100 bug reports randomly selected from the total 1,139 in the dataset (we excluded 141 of them as explained in Section \ref{sec:results}). Each developer analyzed a different set of bug reports and was asked to categorize bugs according to the taxonomy of bug root causes we previously built. During this step, the developers were allowed to either consult the taxonomy (provided in PDF format and containing a description of the bug categories in our taxonomy similar to the one we discuss in Section~\ref{sec:rq1}) or assign new categories if needed.
		
		Once the task was completed, the developers sent back the file annotated with their categorization. Moreover, we gathered comments on the taxonomy and the classification task. As a result, all the participants found the taxonomy \emph{clear} and \emph{complete}: as a proof of that, the tags they assigned were exactly the same as the ones assigned during the phase of taxonomy building. 
	
		\subsection{\revised{\textbf{RQ$_2$}. Characterizing Different Bug Types}}\label{sec:frequency}
			\revised{In the context of this research question, we aimed at providing a characterization of the different bug types discovered in \textbf{RQ$_1$}. More specifically, we took into account three different perspectives such as frequency, relevant topics discussed, and time needed to fix each bug type. In the following subsections, we report the methodology applied to address those perspectives.}
			
			\smallskip
			\revised{\textbf{Frequency Analysis.} To study this perspective, we analyzed how frequently each category of bug root cause in our taxonomy appears. We computed the frequency each bug report was assigned to a certain root cause during the iterative content analysis. It is worth noting that in our study a bug could not be assigned to multiple categories because of the granularity of the taxonomy proposed: we preferred, indeed, working at a level that allows its generalization over software systems having different scope and characteristics. In Section \ref{sec:results} we present and discuss bar plots showing the frequency of each category of root cause in the taxonomy.}
	
			\smallskip
			\revised{\textbf{Topics Analysis.} With this second investigation, we aimed at understanding what are the popular topics discussed within bug reports of different nature. To perform such an analysis, we exploited a well-known topic modeling approach called Latent Dirichlet Allocation (LDA) \cite{blei2003latent,hecking2018topic}. This is a topic-based clustering technique, which can be effectively used to cluster documents in the topics space using the similarities between their topics distributions \cite{wei2006lda}. Specifically, for each bug category of our taxonomy, we adopted the following steps:}
			
			\begin{enumerate}
			
				\item \revised{First, we extracted all the terms composing the bug reports of a certain category;}
				
				\item \revised{An Information Retrieval (IR) normalization process \cite{BaezaYates} was applied. In particular, as bug reports are written in natural language, we first applied (i) spelling correction, (ii) contractions expansion, (iii) nouns and verbs filtering, and (iv) singularization. Then, terms contained in the bug reports are transformed by applying the following steps: (i) separating composed identifiers using the camel case splitting, which splits words based on underscores, capital letters, and numerical digits; (ii) reducing to lower case letters of extracted words; (iii) removing special characters, programming keywords and common English stop words; (iv) stemming words to their original roots via Porter's stemmer \cite{porter1980algorithm};}
				
				\item \revised{Finally, the preprocessed terms are given as input to the LDA-GA algorithm devised by Panichella \etal \cite{panichella2013effectively}. This is an enhanced version of the standard LDA approach that solves an important problem, namely the setting of the parameter $k$, that is the predefined number of topics to extract. In particular, LDA-GA relies on a genetic algorithm that balances the internal cohesion of topics with the separation among clusters. In this way, it can estimate the ideal number of clusters to generate starting from the input terms \cite{panichella2013effectively}.} 
					
			\end{enumerate}
					
			\revised{In Section \ref{sec:results} we report and discuss the topics given as output by the LDA-GA algorithm.}
			
			\smallskip
			\revised{\textbf{Time-to-fix Analysis.} To investigate such a perspective, we followed the procedure previously adopted by Zhang \etal \cite{zhang2012empirical}. Specifically, we mined a typical bug fixing process where (i) a user defines a bug report, (ii) the bug is assigned to a developer, (iii) the developer works and fixes the bug, (iv) the code is reviewed and tested, and (v) the bug is marked as resolved. Correspondingly, we computed five time intervals:}
			
			\begin{itemize}
				
				\item \revised{\emph{Delay Before Response} (\textsf{DBR}). This is the interval between the moment a bug is reported and the moment it gets the first response from development teams;}
			
				\item \revised{\emph{Delay Before Assigned} (\textsf{DBA}). This measures the Interval between a bug getting the first response and its assignment;}

				\item \revised{\emph{Delay Before Change} (\textsf{DBC}). This is the interval between a bug getting assigned and the developer starting to fix the bug, namely she performs the first commit after the bug has been assigned;}
			
				\item \revised{\emph{Duration of Bug Fixing} (\textsf{DBF}): This measures the interval between the developer starting and finishing the bug fixing, namely the time between the first and last commit before the bug has been marked as solved;}
			
				\item \revised{\emph{Delay After Change} (\textsf{DAC}): This is the interval between the developer finishing the bug fixing and the status of the bug being changed to resolved.}
			
			\end{itemize}
	
			\revised{To compute these metrics, we relied on the evolution of the history of each bug report using the features available from the issue tracker. In particular, we mined (1) the timestamp in which a bug has been opened and that of the first comment for computing \textsf{DBR}; (2) the timestamp of the first comment and the one reporting when a bug report changed its status in \textsf{``assigned''} for \textsf{DBA}; (3) the timestamp of the \textsf{``assigned''} status and that of the first commit of the author involving the buggy artifact for \textsf{DBC}; (4) the timestamp of the first and last commit before the bug is marked as solved for \textsf{DBF}; and (5) the timestamp of the last commit and the one reporting the bug as \textsf{``resolved''} for \textsf{DAC}. For all the metrics, in Section \ref{sec:results} we report descriptive statistics of the delay in terms of hours. It is important to note that, as done in previous work \cite{zhang2012empirical}, we filtered out all bugs whose final resolution was not fixed to ensure that only actually fixed bugs were investigated. It is important to note that the detailed results and script of these analyses are available in the online appendix \cite{appendix}.}		
			
		\subsection{\textbf{RQ$_3$}. Automated Classification of Bug Types}\label{sec:automation}
			Our final research question we focused on assessing the feasibility of a classification model able to classify bug root causes starting from bug reports. We relied on machine learning since this type of approach can automatically learn the features discriminating a certain category, thus simulating the behavior of a human expert \cite{pantic2007human}. As a side effect of this research question, we also pose a baseline against which future approaches aimed at more accurately classifying bug root causes can be compared. 
			The following subsections detail the steps followed when building and validating our root cause classification model. 
	
			\smallskip
			\textbf{Independent and Dependent Variables.} Our goal was to classify the root cause of bugs based on bug report information. We exploited summary messages contained in such reports as \emph{independent variables} of our root cause classification model: our choice was driven by recent findings that showed how in most cases bug report summaries properly describe a bug, thus being a potentially powerful source of information to characterize its root cause \cite{zimmermann2010makes}. Moreover, we did not include the title of a bug report as an independent variable because it might contain noise that potentially limits the classification performance \cite{zhou2016combining}. 
			
			It is important to note that not all the words contained in a summary might actually be representative and useful to characterize the root cause of a bug. For this reason, we needed to properly preprocess them \cite{chowdhury2003natural}. 
			
			In our context, we adopted the widespread \emph{Term Frequency - Inverse Document Frequency} (\textsc{TF-IDF}) model \cite{salton1988term}, which is a weighting mechanism that determines the relative frequency of words in a specific document (\ie a summary of bug report) compared to the inverse proportion of that word over the entire document corpus (\ie the whole set of bug report summaries in our dataset). This technique measures \emph{how characterizing} a given word is in a bug report summary: for instance, articles and prepositions tend to have a lower \textsc{TF-IDF} since they generally appear in more documents than words used to describe specific actions~\cite{salton1988term}. 
			Formally, let $C$ be the collection of all the bug report summaries in our dataset, let $w$ be a word, and let $c \in C$ be a single bug report summary, the \textsc{TF-IDF} algorithm computes the relevance of $w$ in $c$ as:
			\begin{equation}
			relevance(w,c) = f_{w,c} \cdot \log (|C|/f_{w,C})
			\end{equation}
			
			where $f_{w,c}$ equals the number of times $w$ appears in $c$, $|C|$ is the size of the corpus, and $f_{w, C}$ is equal to the number of documents in which $w$ appears. The weighted words given as output from \textsc{TF-IDF} represent the independent variables for the classification model. It is important to note that the choice of \textsc{TF-IDF} was driven by experimental results: specifically, we also analyzed the accuracy of models built using more sophisticated techniques such as \textsc{Word2Vec} \cite{goldberg2014word2vec} and \textsc{Doc2Vec} \cite{le2014distributed}. As a result, we observed that the use of \textsc{TF-IDF} led to an improvement of F-Measure up to 13\%. Therefore, we focus on \textsc{TF-IDF} in the remainder of the paper.
			\smallskip 
	
			As for \emph{dependent variable}, it was represented by the bug root causes present in our taxonomy. 
			
			\smallskip
	
			\textbf{Classifiers.} \revised{With the aim of providing a wider overview of the performance achieved by different classifiers,  we experimented with classifiers previously used for prediction purposes by the research community, \ie (i) \textsc{Naive Bayes}, (ii) \textsc{Support Vector Machines} (\textsc{SVM}), (iii) \textsc{Logistic Regression}, and (iv) \textsc{Random Forest}. These classifiers have different characteristics and different advantages/drawbacks in terms of execution speed and over-fitting \cite{nasrabadi2007pattern}. It is important to note that before running the models, we also identified their best configuration using the \textsc{Grid Search} algorithm \cite{bergstra2012random}. Such an algorithm represents a brute force method to estimate hyperparameters of a machine learning approach. Suppose that a certain classifier $C$ has $k$ parameters, and each of them has $N$ possible values. A grid search basically considers a Cartesian product $f_{|_{k \times N}}$ of these possible values and tests all of them. We selected this algorithm because recent work in the area of machine learning has shown that \textsc{Grid Search} is among the most effective methods to configure machine learning algorithms \cite{bergstra2012random}.}
			
			\revised{After the experimental analysis, we found that \textsc{Logistic Regression} provided the best performance. For this reason, in Section~\ref{sec:results} we only report the findings achieved using this classifier. A complete overview of the performance of the other models built with different classifiers is available in our online appendix~\cite{appendix}.}
	
			\smallskip
	
			\textbf{Validation Strategy.} To validate the model, we adopted 10-fold cross validation~\cite{stone1974cross}. It splits the data into ten folds of equal size applying a stratified sampling (\ie each fold has the same proportion of each bug type). One fold is used as a test set, while the remaining ones as a training set. The process is repeated 100 times, using each time a different fold as a test set. Given that the distribution of the dependent variable is not uniform (see more in Section \ref{sec:rq2}), we took into account the problem of training data imbalance \cite{chawla2002smote}. This may appear when the number of data available in the training set for a certain class (e.g., the number of bugs having a certain root cause) is far less than the amount of data available for another class (e.g., the number of bugs having another root cause). More specifically, we applied the Synthetic Minority Over-sampling Technique (\textsc{SMOTE}) proposed by Chawla \etal \cite{chawla2002smote} to make the training set uniform with respect to the root causes available in the defined taxonomy. Since this approach can be run once per time to over-sample a certain minority class, we repeated the over-sampling until all the classes considered have a similar number of instances.
			
			Finally, to cope with the randomness arising from using different data splits \cite{refaeilzadeh2009cross}, we repeated the 10-fold cross validation 100 times, as suggested in previous work \cite{hall2011developing}. We then evaluated the mean accuracy achieved over the runs \cite{stone1974cross}.
			
			To measure the performance of our classification, we first computed two well-known metrics such as \emph{precision} and \emph{recall}~\cite{BaezaYates}, which are defined as follow:
				\begin{equation}
					\mathit{precision} = {{TP} \over {TP+FP}}
					\qquad
					\mathit{recall} = {{TP} \over {TP+TN}}
				\end{equation}
				
				where \emph{TP} is the number of true positives, \emph{TN} the number of true negatives, and \emph{FP} the number of false positives. 
				In the second place, to have a unique value representing the goodness of the model, we compute the F-Measure, \ie the harmonic mean of precision and recall:
				\begin{equation}
					F\mbox{-}Measure = 2 * \frac{precision*recall}{precision+recall}
				\end{equation}
				
				Moreover, we considered two further indicators. The first one is the Area Under the ROC Curve (AUC-ROC) metric. This measure quantifies the overall ability of the classification model to discriminate between the different categories. The closer the AUC-ROC to 1, the higher the ability of the model. In contrast, the closer the AUC-ROC to 0.5, the lower the accuracy of the model.
				Secondly, we computed the Matthews Correlation Coefficient (MCC) \cite{baldi2000assessing}, a regression coefficient that combines all four quadrants of a confusion matrix, thus also considering true negatives: 
				\begin{equation}
					MCC = \frac{(TP*TN) -(FP*FN)}{\sqrt{(TP+FP)(TP+FN)(TN+FP)(TN+FN)}}
				\end{equation}
				where \emph{TP}, \emph{TN}, and \emph{FP} represent the number of (i) true positives, (ii) true negatives, and (iii) false positives, respectively, while \emph{FN} is the number of false negatives.
				Its value ranges between -1 and +1. A coefficient equal to +1 indicates a perfect prediction; 0 suggests that the model is no better than a random one; and -1 indicates total disagreement between prediction and observation.
				

\section{Analysis of the Results}
\label{sec:results}

In this section, we report the results of our study, discussing each research question independently. 
	
	\subsection{\textbf{RQ$_1$}. Taxonomy of Bug Root Cause}
	\label{sec:rq1}
	The manual analysis of the 1,280 bug reports led to the creation of the taxonomy of 9 bug root causes, described in the next subsections. At the same time, we had to discard 141 bug reports for two reasons. In particular, 18 of them---all found in \textsc{Mozilla}---were related to bug reports listing multiple bugs to solve before the release of a new version of the system: from a practical point of view, they represent a to-do list rather than accurate bug reports. For this reason, we decided to exclude them as we could not identify a specific category to which to assign them. On the other hand, 123 bug reports could be considered as false positives due to proposals for improvement or suggestions on how to fix existing bugs: also in this case, we did not consider these suitable for the scope of our study. To some extent, the latter category of false positives highlights how the use of a fully automated filtering technique like the one proposed by Herzig \etal~\cite{herzig2013s} (used in the context selection phase to gather bug reports actually reporting observed malfunctions) is not enough to discard misclassified bugs, \ie the results of such tools must always be double-checked to avoid imprecisions. At the end of this process, the final number of bug reports classified was 1,139.
	In the following, we explain each category of bug root cause in our taxonomy, reporting an example for each of them. Given the excessive length of the bug reports analyzed, we do not report the entire summary in the examples but we highlight the main parts that allow the reader to understand the problem and why we marked it as belonging to a certain root cause category.
	
	\medskip
	
%

	\noindent\textbf{A. Configuration issue.}
	The first category regards bugs concerned with building configuration files. Most of them are related to problems caused by (i) external libraries that should be updated or fixed and (ii) wrong directory or file paths in \texttt{xml} or \texttt{manifest} artifacts. As an example, the bug report shown below falls under this category because it is mainly related to a wrong usage of external dependencies that cause issues in the web model of the application.

	\begin{center}	
	\begin{commitbox}
		\begin{flushleft}
			Example summary.
		\end{flushleft}	
		\emph{``JEE5 Web model does not update on changes in web.xml''}
		
		[Eclipse-WTP Java EE Tools] - Bug report: \texttt{190198}
	\end{commitbox}	 
	\end{center}

	\noindent\textbf{B. Network issue.}
	This category is related to bugs having as root cause connection or server issues, due to network problems, unexpected server shutdowns, or communication protocols that are not properly used within the source code. For instance, in the following, we show an example where a developer reports a newly introduced bug due to a missing recording of the network traffic of the end-users of the project.
	
	\begin{center}	
		\begin{commitbox}
			\begin{flushleft}
				Example summary.
			\end{flushleft}	
			\textit{``During a recent reorganization of code a couple of weeks ago, SSL recording no longer works''}
			
			[Eclipse-z\_Archived] - Bug report: \texttt{62674}
		\end{commitbox}	 
	\end{center}

	\noindent\textbf{C. Database-related issue.}
	This category collects bugs that report problems with the connection between the main application and a database. For example, this type of bug report describes issues related to failed queries or connection, such as the case shown below where the developer reports a \texttt{connection stop} during the loading of a Java Servlet.
	
	\begin{center}	
		\begin{commitbox}
			\begin{flushleft}
				Example summary.
			\end{flushleft}	
			\textit{``Database connection stops action servlet from loading''}
			
			[Apache Struts] - Bug report: \texttt{STR-26}
		\end{commitbox}	 
	\end{center}
	
	\noindent\textbf{D. GUI-related issue.}
	This category refers to the possible bugs occurring within the Graphical User Interface (GUI) of a software project. It includes issues referring to (i) stylistic errors, \ie screen layouts, elements colors and padding, text box appearance, and buttons, as well as (ii) unexpected failures appearing to the users in form of unusual error messages. In the example below, a developer reports a problem that arises because s/he does not see the actual text when s/he types in an input field.
	
		\begin{center}	
		\begin{commitbox}
			\begin{flushleft}
				Example summary.
			\end{flushleft}
			\textit{``Text when typing in input box is not viewable.''}
			
			[Mozilla-Tech Evangelism Graveyard] - Bug report: \texttt{152059}
		\end{commitbox}	 
		\end{center}

	\noindent\textbf{E. Performance issue.} 
	This category collects bugs that report performance issues, including memory overuse, energy leaks, and methods causing endless loops. An example is shown below, and reports a problem raised in the \textsc{Mozilla} project where developers face a performance bug due to the difficulties in loading an external file.
	
	\begin{center}	
		\begin{commitbox}
			\begin{flushleft}
				Example summary.
			\end{flushleft}
			\textit{``Loading a large script in the Rhino debugger results in an endless loop (100\% CPU utilization)''} 
			
			[Mozilla-Core] - Bug report: \texttt{206561}
		\end{commitbox}	 
	\end{center}

	\noindent\textbf{F. Permission/Deprecation issue.} 
	Bugs in this category are related to two main causes: on the one hand, they are due to the presence, modification, or removal of deprecated method calls or APIs; on the other hand, problems related to unused API permissions are included. To better illustrate this category, in the following we provide an example for each of the causes that can fall into this category. The first involves a bug appearing in the case of an unexpected behavior when the method of an external API is called. The second mentions a bug that appears through malformed communication with an API. 
	
	\begin{center}	
		\begin{commitbox}
			\begin{flushleft}
				Example summary.
			\end{flushleft}
			\textit{``setTrackModification(boolean) not deprecated; but does not work''}
			
			[Eclipse-EMF] - Bug report: \texttt{80110}
		\end{commitbox}	

		\smallskip
		\begin{commitbox}
			\begin{flushleft}
				Example summary.
			\end{flushleft}
			\textit{``Access violation in DOMServices::getNamespaceForPrefix (DOMServices.cpp:759)''} 
			
			[Apache-XalanC] - Bug report: \texttt{XALANC-55}
		\end{commitbox}	
	\end{center}
	
	\noindent\textbf{G. Security issue.}
	Vulnerability and other security-related problems are included in this category. These types of bugs usually refer to reload certain parameters and removal of unused permissions that might decrease the overall reliability of the system. An example is the one appearing in the \textsc{Apache Lenya} project, where the \texttt{Cocoon} framework was temporarily stopped because of a potential vulnerability discovered by a developer.
	
		\begin{center}	
			\begin{commitbox}
				\begin{flushleft}
					Example summary.
				\end{flushleft}
				\textit{``Disable cocoon reload parameter for security reasons''} 
				
				[Apache-Lenya] - Bug report: \texttt{37631}
			\end{commitbox}	 
		\end{center}

	\noindent\textbf{\revised{H. Program Anomaly Issue.}}
	\revised{Bugs introduced by developers when enhancing existing source code, and that are concerned with specific circumstances such as exceptions, problems with return values, and unexpected crashes due to issues in the logic (rather than, e.g., the GUI) of the program. It is important to note that bugs due to wrong SQL statements do not belong to this category but are classified as database-related issues because they conceptually relate to issues in the communications between the application and an external database, rather than characterizing issues arising within the application. It is also worth noting that in these bug reports developers tend to include entire portions of source code, so that the discussion around a possible fix can be accelerated. An example is shown below and reports a problem that a developer has when loading a resource.}

	\begin{center}	
	\begin{commitbox} 
		\begin{flushleft}
			Example summary.
		\end{flushleft}
		\textit{``Program terminates prematurely before all execution events are loaded in the model"} 
		
		[Eclipse-z\_Archived] - Bug report: \texttt{92067}
	\end{commitbox}
	\end{center}

	\noindent\textbf{I. Test Code-related issue.} 
	The last category is concerned with bugs appearing in test code. Looking at bug reports in this category, we observed that they usually report problems due to (i) running, fixing, or updating test cases, (ii) intermittent tests, and (iii) the inability of a test to find de-localized bugs. As an example, the bug report below reports on a problem occurred because of a wrong usage of mocking.
	
	\begin{center}	
		\begin{commitbox}
			\begin{flushleft}
				Example summary.
			\end{flushleft}
			\textit{``[the test] makes mochitest-plain time out when the HTML5 parser is enabled"} 
			
			[Mozilla-Core] - Bug report: \texttt{92067}
		\end{commitbox}
	\end{center}

	\subsection{\textbf{RQ$_2$}. The Characteristics of Different Bug Types}
	\label{sec:rq2}
	
	\revised{After we had categorized and described the taxonomy, we focused on determining the characteristics of the different bug types discovered. For the sake of comprehensibility, in this section, we individually discuss the results achieved for each considered aspect, \ie frequency, relevant topics, and time-to-fix process.}
	
	\medskip	
	\revised{\textbf{Frequency Analysis.} With this first perspective, we aimed at studying how prevalent each root cause is in our dataset.}
	
	\begin{figure}[h]
		\centering
		\includegraphics[width=1\linewidth]{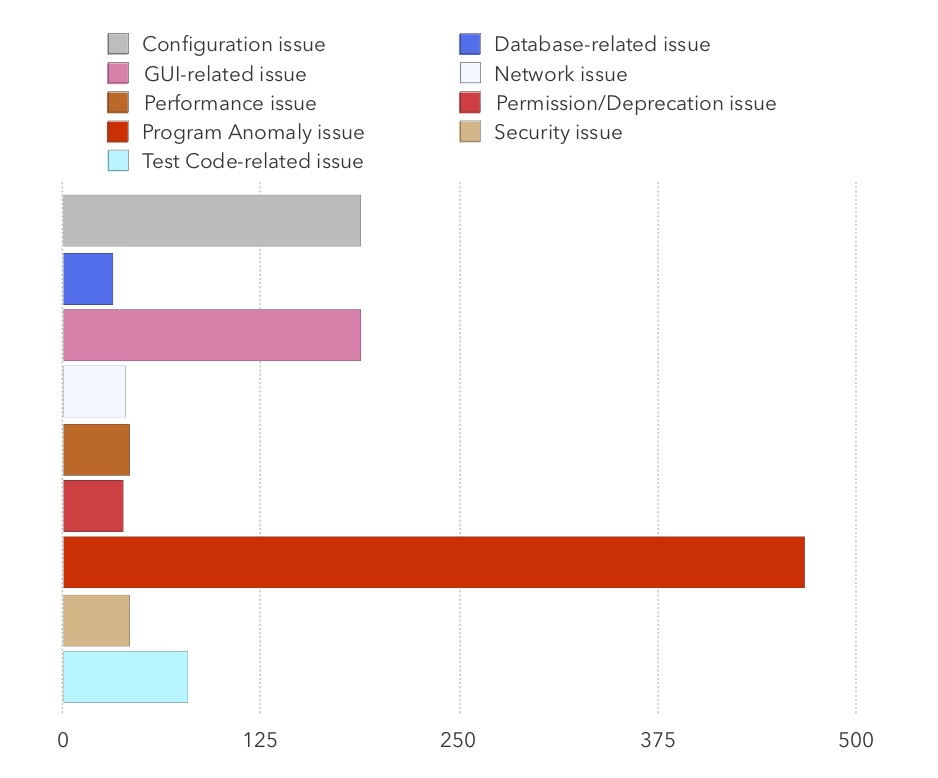}
		\caption{\textbf{RQ$_2$} - Frequency of each category of bug root cause.}
		 \label{fig:frequencies}
		\label{fig:bugReportExample}
	\end{figure}
	
	\begin{table*}[!h]
		\centering
		\caption{\revised{\textbf{RQ$_2$} - Relevant topics of each category of bug root cause.}}
		\label{tab:lda}
		\begin{tabular}{l|c|c|c|c|c}\hline
			\textbf{Categories} & \textbf{Topic 1} & \textbf{Topic 2} & \textbf{Topic 3} & \textbf{Topic 4} & \textbf{Topic 5} \\\hline
			\rowcolor[HTML]{c9c5c5}  Configuration issue & link & file & build & plugin & jdk\\\hline
			Network issue & server & connection & slow & exchange & - \\\hline
			\rowcolor[HTML]{c9c5c5}  Database-related issue & database & sql & connection & connection & - \\\hline
			GUI-related issue & page & render & select & view & font\\\hline
			\rowcolor[HTML]{c9c5c5} Perfomance issue & thread & infinite & loop & memory & - \\\hline
			Permission/Deprecation issue & deprecated & plugin  & goal & - & - \\\hline
			\rowcolor[HTML]{c9c5c5}  Security issue & security & xml & packageaccess & vulnerable & - \\\hline
			Program Anomaly issue& error& file& crash& exception & - \\\hline
			\rowcolor[HTML]{c9c5c5}  Test Code-related issue & fail & test & retry & - & - \\                 
		\end{tabular}
	\end{table*}

	Figure \ref{fig:frequencies} shows the diffusion of root causes extracted from the 1,139 analyzed bug reports. As depicted, the most frequent one is the \emph{Functional Issue}, which covers almost half of the entire dataset (\ie 41,3\%). This was somehow expected as a result: indeed, it is reasonable to believe that most of the problems raised are related to developers actively implementing new features or enhancing existing ones. 
	Our findings confirm previous work \cite{tan2014bug,aranda2009secret} on the wide diffusion of bugs introduced while developers are busy with the implementation of new code or when dealing with exception handling.
	
	\emph{GUI-related} problems are widely present in the bug reports analyzed (17\% of the total number of issues in the dataset). Nowadays, GUIs are becoming a major component of many software systems because they shape the interaction with the end-user. As such, they can evolve and become more complex, thus attracting as many bugs as the codebase \cite{memon2002gui}. This result somehow confirms the findings reported by Tan \etal \cite{tan2014bug}, who also discovered that GUI-related issues are highly popular in modern software systems.
	
	The third most popular root cause is the \emph{Configuration issue} one, as 16\% of the bug reports referred to this root cause. Since the use of external libraries and APIs is growing fast \cite{robbes2012developers,mileva2009mining,salza2018developers}, bugs related to how an application communicates or interacts with external components are becoming more frequent. Moreover, McDonnell \etal \cite{mcdonnell2013empirical} recently showed that the API adaptation code tends to be more bug-prone, possibly increasing the chances of such category of bugs. At the same time, it is also worth noting that some recent findings \cite{bezemer2017empirical} also reported that issues with configuration files (\eg the presence of unspecified dependencies \cite{bezemer2017empirical}) represent a serious issue for developers, which might lead them to introduce more bugs. 
	
	After these first three root causes, we discovered that 7\% of the bug reports in our dataset referred to \emph{test code bugs}. Using another experimental setting, and observing the relative diffusion of this root cause, we confirm the results of Vahabzadeh \etal \cite{vahabzadeh2015empirical}, who showed that the distribution of bugs in test code does not variate too much with respect to that of production code. Our findings are also in line with what is reported in recent studies on the increasing number of test-related issues \cite{luo2014empirical,palomba2017does,bell2018deflaker,DBLP:journals/ese/ZaidmanRDD11,beller2017oops,DBLP:conf/icst/ZaidmanRDD08}.
	
	Performance issues comprise 4\% of the total number of issues. This result confirms the observation from Tan \etal~\cite{tan2014bug}. Indeed, they discovered that bugs related to performance are much less frequent than functional bugs and that their number usually decreases over the evolution of a project. A likely explanation for the relatively low diffusion of this root cause is that the developers involved in the software ecosystems considered in the study often use performance leak detection tools during the development. For instance, the \textsc{Mozilla} guidelines\footnote{\url{https://developer.mozilla.org/en-US/docs/Mozilla/Performance}} highly recommend the use of such tools to limit the introduction of performance leaks in the project as much as possible.
	
	Other specific root causes such as \emph{Network}, \emph{Security}, and \emph{Permission/Deprecation} appear to be less diffused over the considered dataset, \ie they are the cause of $\approx4\%$ reported bugs. 
	Interestingly, our findings related to security-related bugs are not in line with those reported in the study by Tan \etal \cite{tan2014bug}. Indeed, while they found that this root cause is widespread in practice, we could only find a limited number of bug reports actually referring to security problems. 
	Finally, the least spread root cause is \emph{Database-related}, that arises in 3\% of the cases, confirming that such bugs represent a niche of the actual issues occurring in real software systems \cite{schroter2006if}: in this regard, it is worth remarking that replications of our study targeting database-intensive applications would be beneficial to further verify our finding.
	
	To broaden the scope of the discussion, we noticed that the diffusion of the root causes discussed so far is independent from the type of system considered. Indeed, we observed similar distributions over all three ecosystems analyzed, meaning that the same bug categories might basically be found in any software project. This supports our next step: the creation of an automated solution to classify the root cause of bugs, something which could be immediately adopted for improving the diagnosis of bugs. 
		
	\medskip	
	\revised{\textbf{Topics Analysis.} Table \ref{tab:lda} reports the results achieved when applying the LDA-GA algorithm over the bug reports of each category of bug root cause present in our taxonomy. It is important to note that LDA-GA found up to five different clusters that describe the topics characterizing each bug type; a \textsf{`-'} symbol is put in the table in case LDA-GA did not identify more topics for a certain bug type. From a general point of view, we can observe that there is almost zero overlap among the terms describing each bug root cause: on the one hand, we notice that all the topics extracted for each category are strictly related to the description of the categories discussed above (\eg the word \textsf{``test''} describes test-related issues); on the other hand, the lack of overlap is a symptom of a good systematic process of categorization of the bug reports.}
	
	\revised{Going more in depth, the topics extracted for the configuration issue category are very much linked to problems appearing in configuration files and concerned with build issues  (\textsf{``build''} , \textsf{``file''}, \textsf{``jdk''}), caused by wrong file paths (\ie \textsf{``link''}) or external components that should be updated (\ie \textsf{``plugin''}).
	A similar discussion can be delineated in the case of network issues. In particular, based on the bug reports belonging to this category, we found words such as \textsf{``server''} and \textsf{``connection''} that represent topics strictly related to network information, together with words reporting the likely causes of these issues, \ie \textsf{``slow''} connection or problems due to the \textsf{``exchange''} of data over the network.}
	
	\revised{In the context of database-related issues, our findings provide two main observations. The words contained in these bug reports contain clear references to problems occurring with databases, like \textsf{``database''}, \textsf{``SQL''}, or \textsf{``connection''}. At the same time, it is worth noting that the word \textsf{``connection''} occurs twice and, more importantly, is in overlap with a word appearing in network-related bug reports. On the one hand, it is important to note that LDA analysis can generate multiple topics having the same discriminant word \cite{blei2003latent}: indeed, each document (\ie bug reports, in our case) is viewed as a mixture of various topics. That is, for each document LDA-GA assigns the probability of the bug report belonging to each topic. The probability sums to 1: hence, for every word-topic combination there is a probability assigned and it is possible that a single word has the highest probability in multiple topics \cite{blei2003latent}. From a practical perspective, this may indicate that problems with the database connection can be the principal cause of this type of bugs. As for the overlap between network and database issues, this is somehow expected: both the types of bugs have to deal with connection problems of different nature. This might possibly create noise for automated solutions aiming at discriminating different bug types.}
	\revised{Regarding the GUI-related issues, we find that the topics are represented by words clearly related to a GUI interface of a software project \ie \textsf{``page''}, \textsf{``render''}, \textsf{``select''}, \textsf{``view''}, and \textsf{``font''}. For instance, they could concern problems with the rendering of a certain \textsf{``font''} or an entire page; in any case, these are problems with the visualization of the interface of a system.}
	\revised{As for performance-related issues, the topics extracted faithfully represent problems connected with excessive memory consumption; indeed, words such as \textsf{``thread''}, \textsf{``infinite''}, \textsf{``loop''}, and \textsf{``memory''} are the four topics that most frequently appear and that better describe those bug reports.}
	\revised{On the other side, in the category related to security issues we found topics linked to problems of \textsf{``package access''}, but also to \textsf{``vulnerable''} components that may lead to \textsf{``security''} problems. Regarding the topic \textsf{``XML''}, it is important to note that there are a number of security issues involving the configuration of XML parsers and how they interact with the document structure \cite{moradian2006possible,lowis2011vulnerability}. For example, let us consider the validation against untrusted external \textsf{DTD}s (Document Type Declaration) files. The \textsf{DTD} of an XML document is one way to define the valid structure of the document, \ie the rules that specify which elements and values are allowed in the declaration. A security problem may arise in case the server's XML parser accepts an arbitrary external \textsf{DTD} URL and attempts to download the DTD and validate the XML document against it. In this case, an attacker could input any URL and execute a Server Side Request Forgery (SSRF) attack where the attacker forces the server to make a request to the target URL.}
		
	\revised{When considering program anomalies, we noticed that the topics extracted are strictly connected with the description of the category given in the context of \textbf{RQ$_1$}. Indeed, topics such as \textsf{``error''}, \textsf{``file''}, \textsf{``patch''}, \textsf{``crash''}, and \emph{exception} are concerned with problems caused by issues in the logic of the program (\eg a wrong return value or an exception). Finally, permission/deprecation and test code-related issues follow the same discussion: all the words extracted by LDA-GA have clearly something to do with their nature: as an example, the word \textsf{``retry''} appearing in tests is connected with a JUnit annotation (\texttt{@Retry}) that highlights the presence of some form of test flakiness, \ie unreliable tests that exhibit a pass and fail behavior with the same code \cite{palomba2017does}.}	

	\revised{All in all, our results indicate that the words characterizing the identified bug root causes are pretty disjoint from each other. As a consequence, it is reasonable to use the words occurring within bug reports to classify them according to the defined taxonomy. This is a clear motivation for adopting a natural language-based machine learning approach like the one proposed in \textbf{RQ$_3$}.}
		
	\begin{figure*}[!h]
		\centering
		\includegraphics[width=1\linewidth]{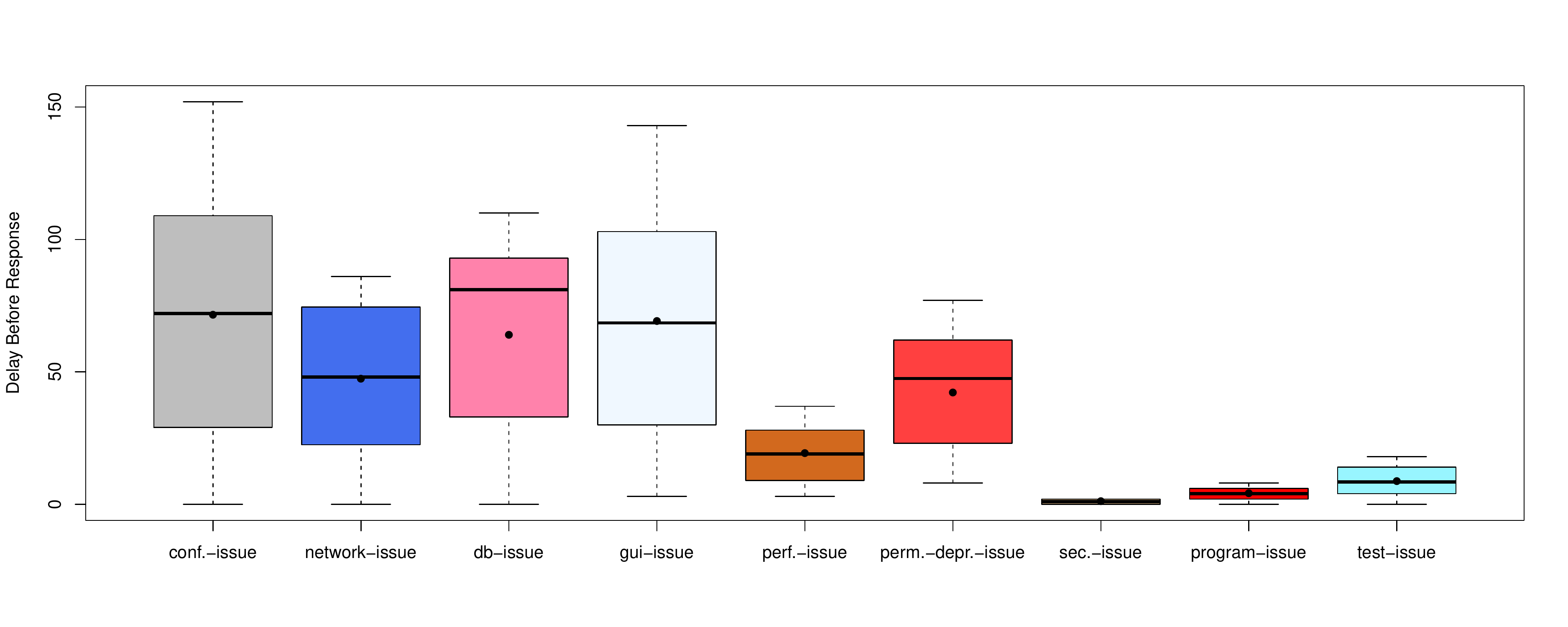}
		\caption{\revised{\textbf{RQ$_2$} - Box plots reporting the Delay Before Response (\textsf{DBR}) for each identified bug root cause.}}
		\label{fig:DBR}
	\end{figure*}

	\begin{figure*}[!h]
		\centering
		\includegraphics[width=1\linewidth]{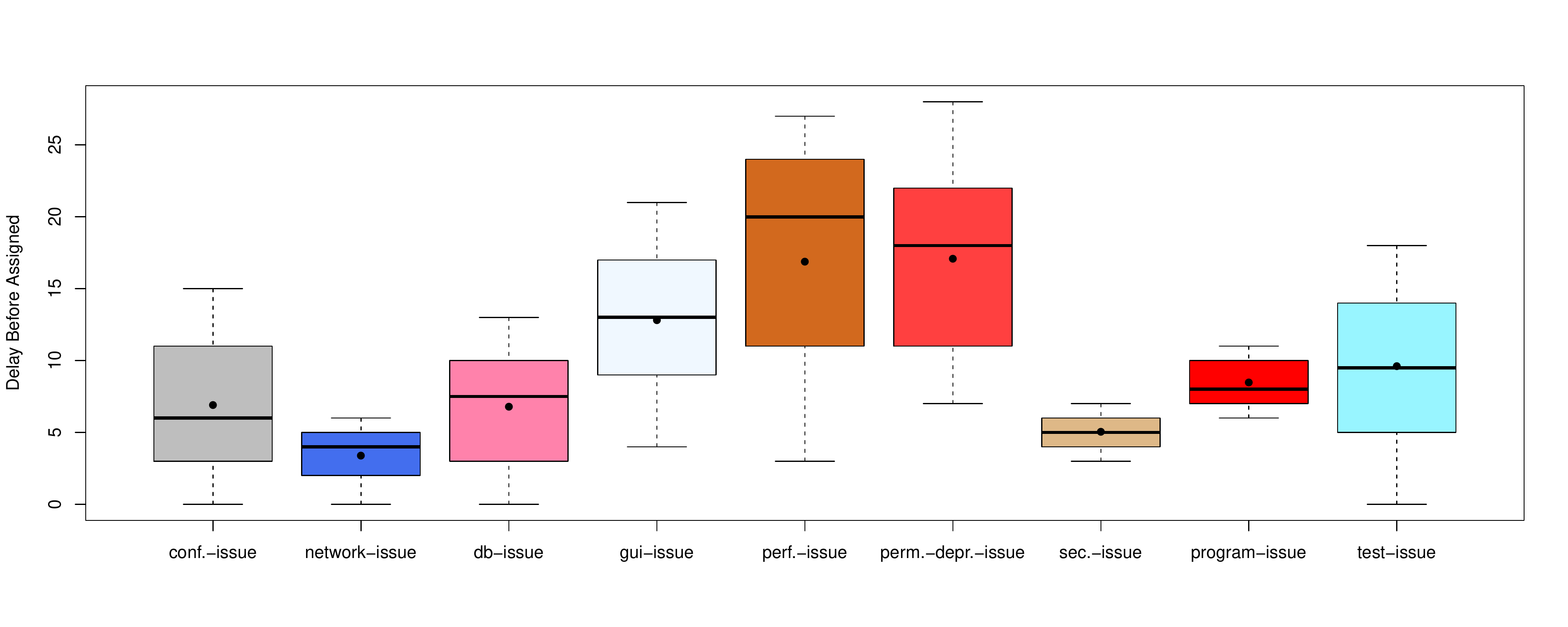}
		\caption{\revised{\textbf{RQ$_2$} - Box plots reporting the Delay Before Assigned (\textsf{DBA}) for each identified bug root cause.}}
		\label{fig:DBA}
	\end{figure*}

	\begin{figure*}[!h]
		\centering
		\includegraphics[width=1\linewidth]{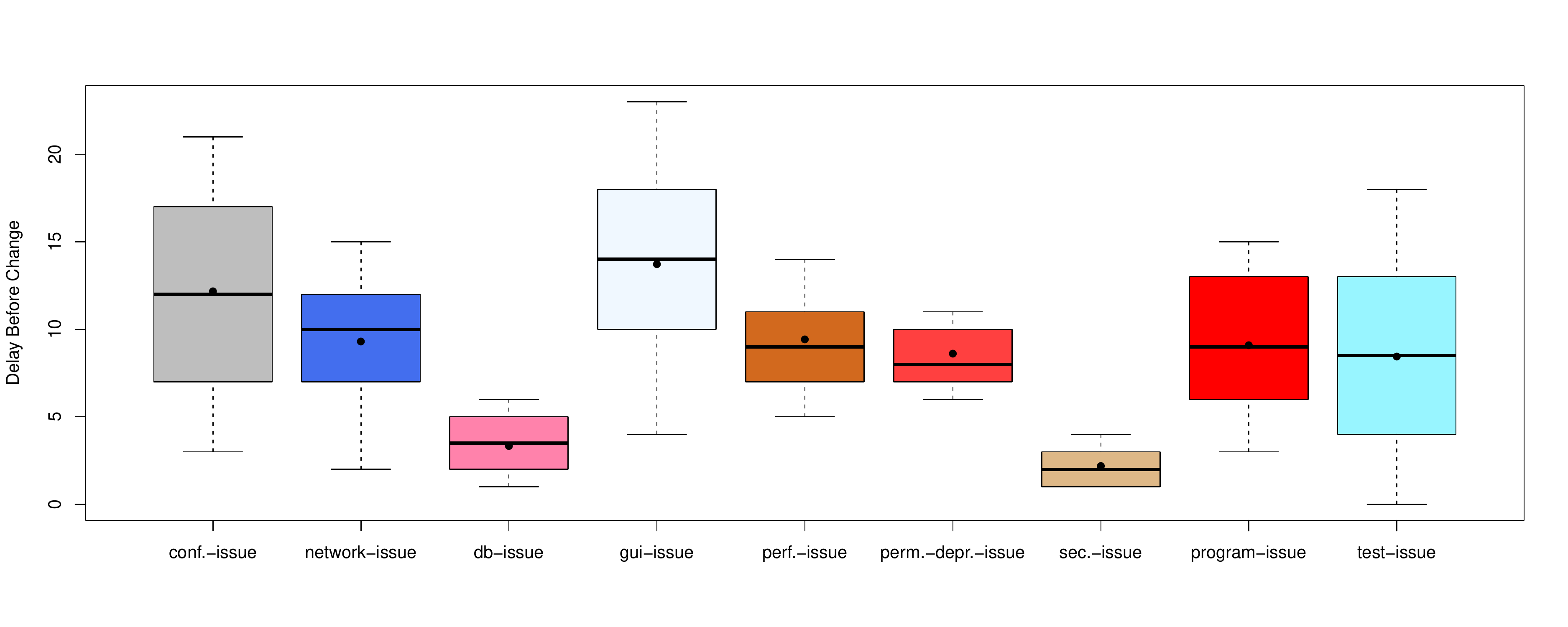}
		\caption{\revised{\textbf{RQ$_2$} - Box plots reporting the Delay Before Change (\textsf{DBC}) for each identified bug root cause.}}
		\label{fig:DBC}
	\end{figure*}	
	
	\begin{figure*}[!h]
		\centering
		\includegraphics[width=1\linewidth]{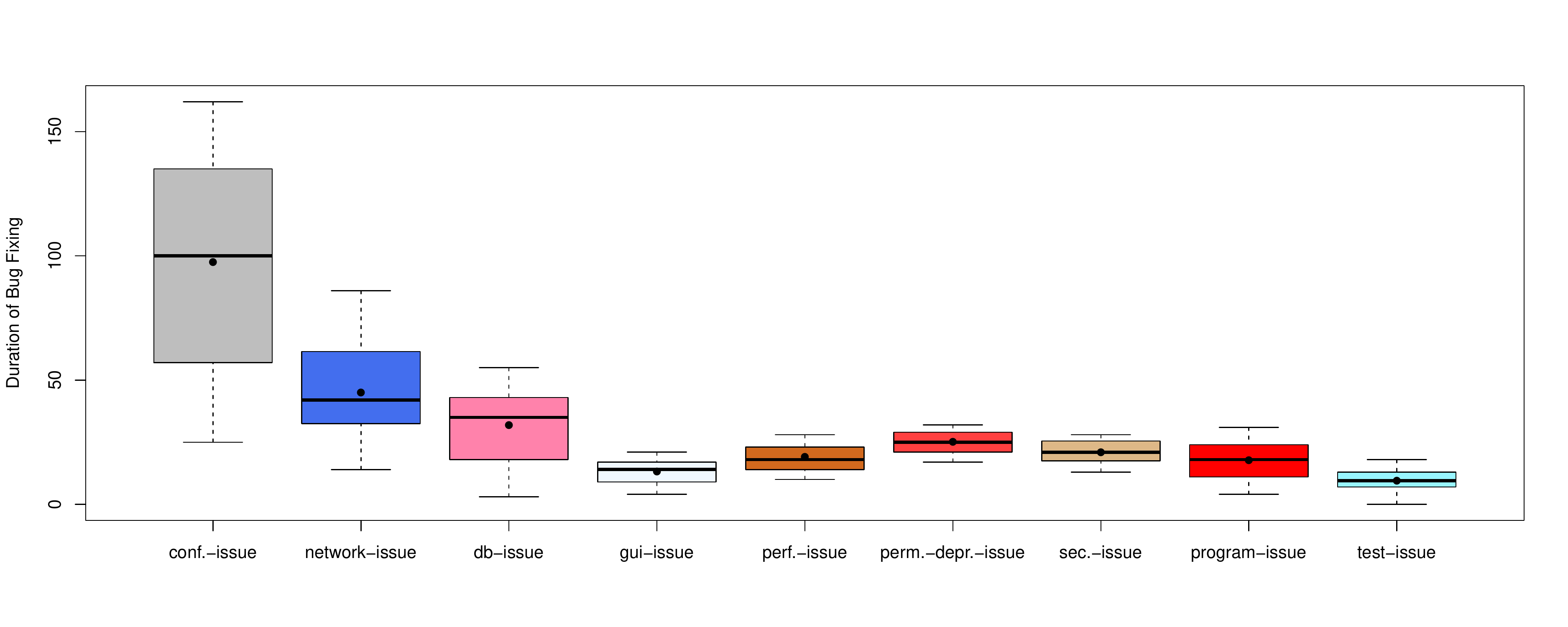}
		\caption{\revised{\textbf{RQ$_2$} - Box plots reporting the Duration of Bug Fixing (\textsf{DBF}) for each identified bug root cause.}}
		\label{fig:DBF}
	\end{figure*}
	
	\begin{figure*}[!h]
		\centering
		\includegraphics[width=1\linewidth]{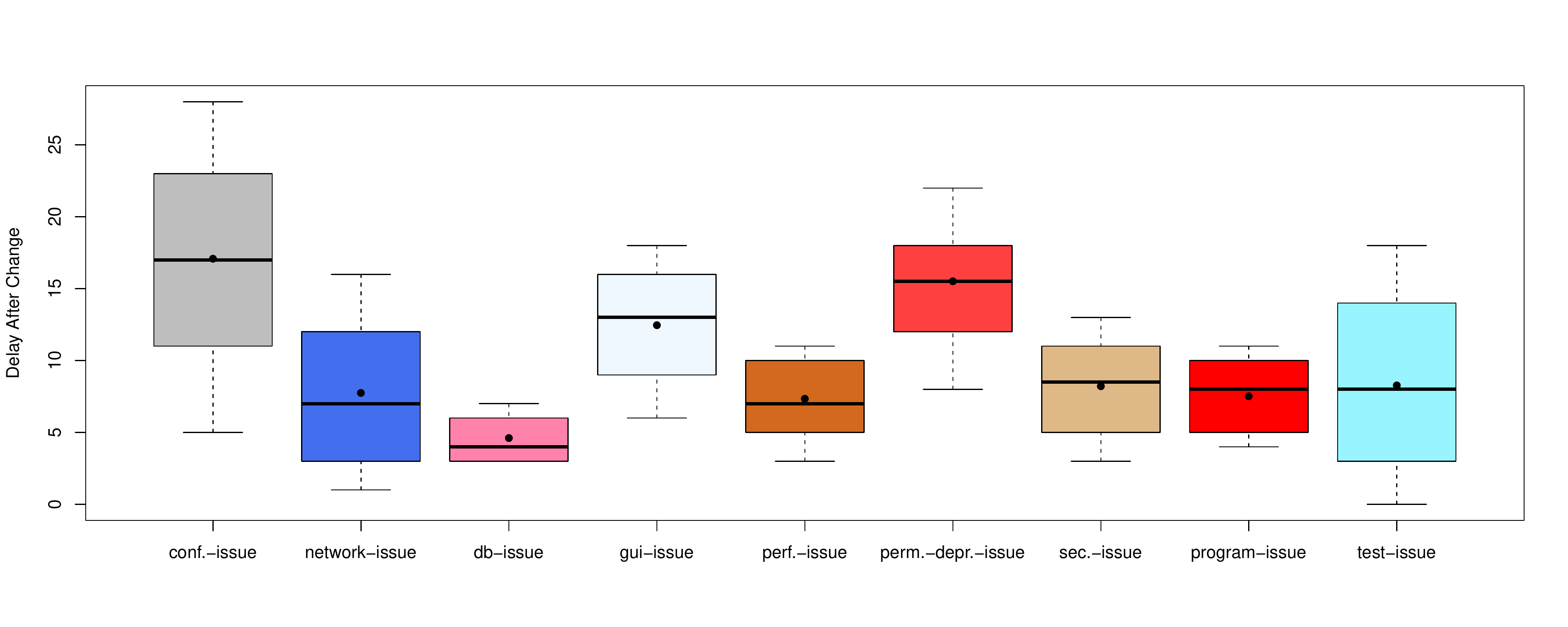}
		\caption{\revised{\textbf{RQ$_2$} - Box plots reporting the Delay After Change (\textsf{DAC}) for each identified bug root cause.}}
		\label{fig:DAC}
	\end{figure*}

	\medskip	
	\revised{\textbf{Time-to-fix Analysis.}} 
	\revised{Figures \ref{fig:DBR}, \ref{fig:DBA}, \ref{fig:DBC}, \ref{fig:DBF}, and \ref{fig:DAC} depict box plots of the five metrics considered to measure the time required from the entire bug fixing process of each bug root cause belonging to the taxonomy, \ie \emph{Delay Before Response}, \emph{Delay Before Assigned}, \emph{Delay Before Change}, \emph{Duration of Bug Fixing}, and \emph{Delay After Change}, respectively. The black dots presented in the figures represent the mean value of each distribution.} 
	
	\revised{From a high-level view, we could first confirm that not all bugs are the same, as each root cause has its own peculiarities in terms of the time required for the entire bug fixing process. Looking more in-depth on the single indicators, the first observation is related to \textsf{DBR}: in this case, we see that security-related issues are those having the smallest delay from reporting to the first response from the development team. This is somehow expected, since security issues have a high harmfulness for the overall reliability of a software system \cite{di2008evolution}. More specifically, both mean and median values are equal to 2, indicating that in a pretty short time window the developers tend to take action after a potential vulnerability is detected. Moreover, the distribution is all around the median, meaning that there is no variability in the time to response among the analyzed security issues: thus, we can claim that independently from the system or other factors such as their frequency of appearance, these issues are seriously taken into account by developers.}
		
	\revised{Also bugs due to program anomalies present a limited time interval between their discovery and the first reaction from developers. Also in this case, the result is quite expected: indeed, this category relates to issues in the inner-working of a program that might potentially have negative consequences on reliability and make the system less appealing for end-users \cite{bavota2015impact,palomba2018crowdsourcing,palomba2017recommending}. The distribution is close to the median, thus highlighting that developers pay immediate attention to these bugs.}
	
	\revised{A more surprising result is the one obtained for test-related issues. Even though they are generally perceived as less important than bugs appearing in production \cite{meyer2014software,tufano2016empirical}, our data shows that developers react pretty fast to their reporting: both mean and median are equal to 7, meaning that the reaction of developers is observed within one week. Also in this case, the distribution is not scattered and, therefore, we can claim that the short-time reaction to test-related issues represents a rule rather than the exception. Likely, this result reflects the ever increasing importance that test cases have in modern software systems, \eg for deciding on whether to integrate pull requests  or build the system \cite{beller2017oops,gousios2015work,DBLP:conf/sigsoft/BellerGPZ15,bellerTSE}.}
		
	\revised{Performance issues have a median delay before response of 12. When comparing this value with those achieved by other types of bugs, we can say that it is pretty low and that, as a consequence, the developers' reaction to this kind of bugs is fast. Our finding confirms previous analyses conducted in the field of performance bug analysis \cite{jovic2011catch}. As for the rest, all the other bug types have much higher distributions, and this likely indicates that developers tend to focus first on issues that directly impact functionalities and reliability of the system.}
		
	\revised{Turning the attention to \textsf{DBA} (Figure \ref{fig:DBA}), we observe that network-related issues are those assigned faster for fixing. If we compare the time required to performance-related issues or permission/deprecation issues to be assigned, we can hypothesize that the observed findings are strictly connected to the difficulty to find good assignees. For instance, a possible interpretation of our results is that, based on the developers' expertise and workload, a certain type of bug is assigned faster than others. While further investigations around this hypothesis would be needed and beneficial to study the phenomenon deeper, we manually investigated the bugs of our dataset to find initial compelling evidence that suggests a relation between time-to-assignment and developer-related factors. As a result, looking at both bug reports and comments, we found 21 cases (out of the total 42) in which the assignment of performance issues has been delayed because of the lack of qualified personnel. For example, let consider the following comment made by a \textsc{Mozilla} developer:}
					
	\begin{center}	
		\begin{quotation}
			\revised{\textit{``I'm reluctant to do such a task, not really and expert... maybe something for E.?"}}
		\end{quotation}
	\end{center}

	\revised{Conversely, in the cases of network-related and security issues, we observed that there exist specific developers that have peculiar expertise on these aspects: this potentially make the assignment faster. Nonetheless, our future research agenda includes a more specific investigation on the factors impacting bug fixing activities; in the context of this paper, we only limit ourselves in reporting that it is possible to observe differences in the way different bug types are treated by developers.}
	
	\revised{As for the \textsf{DBC} (Figure \ref{fig:DBC}), we still observe differences among the discovered bug root causes. Security issues are those that developers start working on faster: as explained above, this is likely due to the importance of these issues for reliability. At the same time, bugs due to database-related problems have a small time interval between assignment and beginning of the actual fixing activities (median=4 hours). Also in this case, it is reasonable to believe that these bugs can cause issues leading end-users not to interact with the system in a proper manner and, therefore, they represent issues that are worth to start fixing quickly. More surprisingly, the fixing process of program anomalies requires a higher number of hours to be started. While more investigations would be needed, we can conjecture that factors like severity and priorities assigned for their resolution have an important impact on how fast developers deal with them.}
		
	\revised{Looking at \textsf{DBF}, namely the duration of bug fixing, we can observe that the differences are less evident than the other metrics. Indeed, the fixing duration of most of the bugs ranges between 2 and 30 hours. This is especially true for program anomalies, GUI and test code-related issues, and security problems. A different discussion is the one for database- and network-related issues: despite them being quickest in terms of \textsf{DBA} and \textsf{DBC}, respectively, their duration is much longer than other bugs. Factors like the complexity of the solution or priority assigned to them might explain such a difference. Overall, however, it seems that developers tend to focus more and more quickly on certain types of bugs, confirming the fact that not all bugs are treated in the same manner.}
	
	\revised{Finally, when considering the \textsf{DAC} reported in Figure \ref{fig:DAC}, we observe that the majority of bug types have a similar delay after that the corresponding patches have been submitted. Most likely, this heavily depends on the processes adopted within the projects to control for the soundness of a patch: for instance, most of the modern projects perform code review activities of all the newly committed code changes, and have standard procedures to assess the validity of the change before integration in the code base \cite{pascarella2018information}. The only exception to this general discussion is related to the configuration-issue, which takes up to 33 hours to be integrated: however, given previous findings in literature \cite{anvik2011reducing,mockus2002two,twidale2005exploring}, we see this as an expected result because configuration-related discussions generally trigger more comments by developers since a change in configuration files might impact the entire software project. As a consequence, they take more time to be actually integrated.
	}
								
	\begin{table}[h]
		\captionsetup{justification=centering}
		\caption{\textbf{RQ$_3$} - Performance (in percentage) achieved by the root cause prediction model. \\P=Precision; R=Recall; F-M=F-Measure; AR=AUC-ROC; MCC=Matthews Correlation Coefficient}
		\label{tab:performance}
		\centering
		\resizebox{1\linewidth}{!}{
			\begin{tabular}{lrrrrr}\hline
				\multirow{2}{*}{Categories} & \multicolumn{5}{c}{Logistic Regression}\\\cline{2-6}
				& P & R & F-M & AR  & MCC\\ \hline
				
				\cellcolor[gray]{0.85}{Configuration issue} & \cellcolor[gray]{0.85}{46} & \cellcolor[gray]{0.85}{52} & \cellcolor[gray]{0.85}{49} & \cellcolor[gray]{0.85}{68} & \cellcolor[gray]{0.85}{66}\\
				
				Database-related issue & 71 & 63 & 67 & 72 & 76 \\
				
				\cellcolor[gray]{0.85}{GUI-related issue} & \cellcolor[gray]{0.85}{61} & \cellcolor[gray]{0.85}{68} & \cellcolor[gray]{0.85}{65} & \cellcolor[gray]{0.85}{77}& \cellcolor[gray]{0.85}{65} \\
				
				Network issue & 36 & 40 & 38 &56 &59\\
				
				\cellcolor[gray]{0.85}{Performance issue} & \cellcolor[gray]{0.85}{67} & \cellcolor[gray]{0.85}{57} & \cellcolor[gray]{0.85}{62} & \cellcolor[gray]{0.85}{65}& \cellcolor[gray]{0.85}{67} \\
				
				Permission/Deprecation issue & 86 & 55 &67 & 69 &74\\
								
				\cellcolor[gray]{0.85}{Program Anomaly issue} & \cellcolor[gray]{0.85}{68} & \cellcolor[gray]{0.85}{65} & \cellcolor[gray]{0.85}{67} & \cellcolor[gray]{0.85}{74} & \cellcolor[gray]{0.85}{68}\\
				
				Security issue & 76 & 74 & 75 & 88 & 85 \\
				
				\cellcolor[gray]{0.85}{Test Code-related issue} & \cellcolor[gray]{0.85}{90} & \cellcolor[gray]{0.85}{70} & \cellcolor[gray]{0.85}{79} & \cellcolor[gray]{0.85}{93} & \cellcolor[gray]{0.85}{88}\\\hline
				
				\cellcolor[gray]{.0}{\textbf{\textcolor{white}{Overall}}} & \cellcolor[gray]{.0}{\textbf{\textcolor{white}{67}}} & \cellcolor[gray]{.0}{\textbf{\textcolor{white}{60}}} & \cellcolor[gray]{.0}{\textbf{\textcolor{white}{64}}} & \cellcolor[gray]{.0}{\textbf{\textcolor{white}{74}}} & \cellcolor[gray]{.0}{\textbf{\textcolor{white}{72}}}\\\hline
				
			\end{tabular}
		}
	\end{table}

	\subsection{\textbf{RQ$_3$}. Automated Classification of Bug Root Causes}
	\label{sec:rq3}
	Table \ref{tab:performance} reports, for each root cause, the mean \emph{precision}, \emph{recall}, \emph{F-measure}, \emph{AUC-ROC}, and \emph{Matthews correlation coefficient} achieved by our root cause prediction model over the 100 runs of the 10-fold cross validation.
	We observed that the F-Measure ranges between 35\% and 77\%, the AUC-ROC between 56\% and 93\%, while the MCC between 59\% to 88\%. Thus, overall, we can claim that the devised prediction model is reasonably accurate in identifying the root cause of a bug by exploiting bug report information. It is important to remark that the model considers the words composing the bug report summary as an independent variable: the model is already able to achieve high performance for most of the categories only taking into account such words, meaning that our initial step toward the automatic classification of bug root causes based on bug report information can be considered successful. Nevertheless, further research on the features that influence the root cause of bugs (\eg structural information of the involved classes) might still improve the performance. We plan to perform a wider analysis of additional features in our future research.
	
	Looking more in-depth into the results, the first interesting observation can be made when analyzing the performance of the model on the \emph{Test Code-related issue} category. In this case, it reaches the highest F-Measure, AUC-ROC, and MCC values (\ie 77\%, 93\%, and 88\%, respectively). Since the model relies on bug report words, the result can be explained by the fact that the terms used by developers in bug reports involving test-related issues are pretty powerful to discriminate this root cause. As a matter of fact, 87\% of the test-related bug reports in our dataset contain terms like ``test'' or ``test suite'', as opposed to bug reports related to different root causes. This means that a textual-based learning model can more easily classify this root cause. For instance, let us consider the bug report number \texttt{358221} available on the \texttt{Lyo} project of the \emph{Eclipse} ecosystem, which reports the following summary:
	
	\begin{center}	
		\begin{commitbox}
			\textit{``Investigate possible test suite bug when ServiceProviderCatalog contains ref to serviceProvider resource''} 
		\end{commitbox}
	\end{center}
	
	Similar observations can be made to explain the results for the \emph{Security issue} category (AUC-ROC=88\%). Also in this case, developers frequently adopt terms like ``security'' or ``vulnerability'' to describe a bug having this root cause. 
	
	Still, categories related to \emph{Functional Issue}, \emph{GUI-related issue}, and \emph{Network issue} can be accurately classified by the model. Specifically, F-Measure values of respectively 67\%, 64\%, and 62\%  are reached. On the one hand, these results confirm that a textual-based approach can be effective in classifying the root cause of bugs. At the same time, our findings eventually reveal that developers follow specific patterns when describing issues related to different categories. 
	
	Turning the attention toward the categories for which the model does not perform very well, there are two main cases to discuss. The first one is related to the \emph{Configuration issue} root cause, which has an F-Measure=48\%. To better understand the possible causes behind this result, we manually analyzed the bug reports belonging to this root cause. Let us consider two cases coming from the \textsc{Apache XalanC} project (bug reports number \texttt{XALANC-44} and \texttt{58288}): 
		
	\begin{center}	
		\begin{commitbox}
			\begin{center}
				\textit{``Could not compile''} 
			\end{center}	
		\end{commitbox}
	\end{center}
	
	\begin{center}	
		\begin{commitbox}
			\textit{``VE hangs; times out; then throws NPE doing pause/reload''} 
		\end{commitbox}
	\end{center}
	
	Looking at these bug reports, we could not immediately understand the root cause they refer to. Indeed, during the taxonomy building phase we could analyze other information like developers' discussions and attachments; however, since our classification model is only based on words composing the summary, sometimes it cannot associate such words to the correct root cause. To some extent, our finding contextualizes the findings by Zimmermann \etal \cite{zimmermann2010makes} on the quality of bug reports, showing it varies depending on the root cause developers have to report. 
	
	\revised{A similar situation arises when considering \emph{Database-related} issues. While in \textbf{RQ$_2$} we discovered that the corresponding bug reports have textual characteristics that might be exploited to identify their root-cause, we also highlighted the presence of overlapping words with other categories that may preclude the correct functioning of the model. As such, this finding indicates once again that the performance of our root cause classification model may be improved by considering further bug report components such as developers' discussions and attachments.}
	
	\revised{To conclude the discussion, it is worth relating the performance of the classification model to the results reported in \textbf{RQ$_2$} on the diffusion of each root cause. Put into this context, the devised model is able to properly predict all the most diffused categories, with the notable exception of \emph{Configuration} issues. As such, we argue that the model can be useful in practice and that more research is needed in order to improve its capabilities in detecting configuration-related problems.}

\section{\revised{Discussion and Implications}}
\label{sec:discussion}

	\revised{Our results highlighted a number of points to be further discussed as well as several practical implications for both practitioners and researchers.}
	
	\medskip
	\revised{\textbf{Discussion.} At the beginning of our investigation, we conjectured that the knowledge of the underlying root cause of bugs could be useful information to exploit to improve bug triaging approaches. Our findings clearly highlighted that bugs are different in nature, have different characteristics with respect to the way developers deal with them, and can be classified with a pretty high accuracy using machine learning models. We argue that this  information can be useful for the bug triaging process for the following three main reasons:}
	
	\begin{itemize}
		
		\item \revised{\textbf{Raising awareness on the decision-making process.} In the first place, through \textbf{RQ$_2$} we discovered that different bugs are subject to a different bug fixing process with respect to both the time they required to be assigned and to be actually fixed and integrated into production. Our automatic classification technique can firstly support developers in the decision-making process, as it can immediately pinpoint the presence of bugs having a nature making them potentially more dangerous than others: as an example, our technique can tag a newly reported bug report as a security issue, raising the \emph{awareness} of developers on the need to take prompt actions, thus possibly speeding up their reaction time, considering its assignment and resolution as well as the in-between activities, \eg pushing the assigned developer(s) to perform the required bug fixing action in a timely manner.} 
		
		\item \revised{\textbf{Comprehending the root cause of bugs.} As a complementary support to awareness, the findings reported in our study have the potential to make developers more focused on the signaled root cause of a reported bug, thus reducing the time required in the understanding of the problem. We hypothesise that this could help reduce the time required to (i) assign a bug to the most-skilled developer and (ii) involve the most-qualified developers in the discussion on how to fix the bug. For instance, the output of the proposed approach would support developers in timely spotting configuration-related issues, that are those having the most delayed fixing process according to our analyses. As a result, community shepherds and developers could use this information to take actions and involve the appropriate set of experienced developers in an effort of finding a solution to fix the newly submitted bug.}
			
		\item \revised{\textbf{Improving Bug Triage.} Finally, the root-cause prediction model proposed in this study can be exploited within existing but triaging approaches to improve their performance. As reported by Shokripour \etal \cite{shokripour2013so}, current approaches can be broadly divided into two sets: activity- and location-based. The former identifies the most-skilled developer to be assigned to a new bug on the basis of her previous activities, namely on the analysis of which bugs she fixed, while the latter takes into account the location of the bug within the source code. More recently, Tian \etal \cite{tian2016learning} proposed a model that combined these two approaches: they considered both developers' previous activities (\eg developer bug fixing frequency) and code locations associated with a bug report as similarity features in order to capture the similarity between a bug report and developers' profile. Nevertheless, all these approaches only consider the developers' perspective, without taking into account the nature of the bug that needs to be fixed. Our model can complement all the existing techniques by complementing the information on the location of a newly reported bug with developers' activities aimed at fixing \emph{specific bug types rather than their merely attitude to resolve bugs}: we envision the definition of novel ensemble approaches and/or search-based algorithms that can exploit the bug root cause together with developers' experience and location of the bug to improve the identification of the most-skilled developer that can fix the bug.} 
				
	\end{itemize}

	\revised{We believe that all the aspects reported above deserve more attention, especially on the basis of the results reported in our study. They are, therefore, part of our future research agenda, which is devoted to the improvement of current bug triaging approaches.}

	\medskip
	\revised{\textbf{Implications.} Besides the discussion points reported above, we see some important implications of our work. More specifically:}
	
	\begin{enumerate}
		
		\item \revised{\textbf{A better understanding of bugs is needed.} In our work, we discovered a number of issues being reported in bug reports: bugs are different from each other, and it would be particularly useful to better study the characteristics of each of them, \eg investigating whether they are introduced differently, with the aim of improving or specializing bug localization approaches and bug prediction models. Moreover, we believe that particular attention should be devoted to the understanding of functional bugs, which are those that appear more frequently in practice. For instance, further studies aimed at decomposing the category in multiple more specific sub-categories or investigating their perceived harmfulness would be beneficial to provide an improved support to developers.}
		
		\smallskip
		\item \revised{\textbf{More research on test code bugs is needed.} Our work revealed that a large number of bugs impact test code. The research community has heavily studied production bugs \cite{ray2016naturalness}, however, only a few studies are available with respect to bugs in test code \cite{luo2014empirical,palomba2017does,bell2018deflaker}. We argue that more research on these bugs can be worthwhile to improve both quality and reliability of test cases. }
		
		\smallskip
		\item \revised{\textbf{Configuration checkers are important.} According to our findings, configuration-related bugs are among the most popular ones. Unfortunately, little is known on this category of bugs \cite{bezemer2017empirical} and there are no tools able to support developers in managing the quality of configuration files. We argue that more research aimed at devising such configuration quality checkers is needed to assist practitioners and avoid the introduction of bugs.}
		
		\smallskip
		\item \revised{\textbf{Establishing whether the role of other bug report features would improve root cause analysis.} While a key results of our work is the good performance of a classification model relying on bug summaries as independent variable, we noticed that in some cases it cannot perform well because words contained in bug reports are not enough to identify the root cause of bugs. On the one hand, studies investigating the linguistic patterns used by developers would be worthwhile to learn how to better classify bug reports; on the other hand, the analysis of the value of other bug report features (\eg developers' discussions) would represent the next step toward an improved support for root cause analysis.}
		
	\end{enumerate}

\section{Threats to Validity}
\label{sec:threats}
In this section, we discuss possible threats affecting our results and how we mitigated them.\smallskip

\subsection{Taxonomy validity} 
To ensure the correctness and completeness of the root causes identified in the taxonomy building phase, we performed an iterative content analysis that allowed us to continuously improve the quality of the taxonomy by merging and splitting categories if needed. Moreover, as an additional validation, we involved $5$ expert industrial developers and asked them to classify a set of 100 bug reports according to the proposed taxonomy. They related the sampled bug reports to the same root causes as those assigned by us during the phase of taxonomy building, thus confirming the completeness and clarity of the identified root causes. 
Nevertheless, we cannot exclude that our analysis missed specific bug reports that hide other root causes.

\subsection{Conclusion Validity}
Threats to conclusion validity refer to the relation between treatment and outcome. \revised{In the context of \textbf{RQ$_2$}, we extracted relevant topics within bug reports referring to different bug root causes using Latent Dirichlet Allocation (LDA) \cite{blei2003latent}. To overcome the problem of configuring the parameter $k$---whose wrong configuration has been shown to bias the interpretation of the results \cite{peng2001lda}---we employed the LDA-GA version of the technique proposed by Panichella \etal \cite{panichella2013effectively}: this is based on a genetic algorithm that is able to exercise the parameter $k$ until an optimal number of clusters is found. Still in \textbf{RQ$_2$}, we investigated the time required for the bug fixing process of different bug root causes by replicating the study of Zhang \etal \cite{zhang2012empirical}, thus taking into account all the metrics they employed to measure the bug fixing process. Nonetheless, it is important to point out that further empirical analyses aimed at understanding the specific reasons behind the observed findings, namely what are the factors that developers take into account when treating different bug types, would be needed: indeed, in our study, we limit ourselves to observing that not all bugs are equal and are indeed treated differently.}
In order to evaluate the root cause prediction model, we measured the performance using a number of different indicators such as precision, recall, F-Measure, AUC-ROC, and MCC, which can provide a wide overview of the model performance. As for the validation methodology, we relied on 10-fold cross validation. While such a strategy has recently been criticized~\cite{Tantithamthavorn:tse2017}, we tackled its main issue, \ie the randomness of the splits, by running it 100 times. Finally, it is worth noting that before running the model, we configured its parameters using the \textsc{Grid Search} algorithm \cite{bergstra2012random}. \revised{Given the nature of the validation strategy adopted, we discussed the overall prediction capabilities of the classification model, while we did not provide the detailed confusion matrix: however, this was not possible in our case because we built 1,000 different confusion matrices due to the 100-times 10-fold cross validation. This would have made the interpretation of the results hard.}

\subsection{External validity} 
Threats in this category mainly concern the generalizability of the results. We conducted this study on a large sample of 1,280 bug reports publicly available on the bug tracking platforms of the considered ecosystems. Such a sample allowed us to get bug reports belonging to 119 different projects. However, we are aware that the proposed taxonomy may differ when considering other systems or closed-source projects. Similarly, the performance of our root cause classification model might be lower/higher on different projects than the ones reported herein. 

\section{Conclusion and Future Directions}
\label{sec:conclusions}

	Not all bugs are the same. Understanding their root causes can be useful for developers during the first and most expensive activity of bug triaging~\cite{akila2015effective}, \ie the diagnosis of the issue the bug report refers to. While several previous works mainly focused on supporting the bug triage activity with respect to the identification of the most qualified developer that should take care of it \cite{murphy2004automatic,javed2012automated}, they basically treat all bugs in the same manner without considering their root cause \cite{zaman2011security}.
	
	\revised{In this paper, we started facing this limitation, by proposing (i) an empirical assessment of the possible root causes behind bugs, (ii) a characterization study of the different root causes identified, and (iii) a classification model able to classify bugs according to their root cause.}
	
	To this aim, we first proposed a novel taxonomy of bug root causes, conducting an iterative content analysis on 1,280 bug reports of 119 software projects belonging to three large ecosystems such as \textsc{Mozilla}, \textsc{Apache}, and \textsc{Eclipse}. 
	\revised{Then, we studied the discovered bug root causes under three different perspectives such as (i) frequency of appearance, (ii) principal topics present in the corresponding bug reports, and (iii) time required to fix them.}
	Finally, we devised a root cause prediction model that classifies bug reports according to the related root cause. We empirically evaluated our root cause classification model by running it against the dataset that came out of the taxonomy building phase, measuring its performance adopting a 100 times 10-fold cross validation methodology in terms of F-Measure, AUC-ROC, and Matthew's Correlation Coefficient (MCC).
	
	The results of the study highlight nine different root causes behind the bugs reported in bug reports, that span across a broad set of issues (\eg GUI-related vs. configuration bugs) and are widespread over the considered ecosystems. 
	\revised{We observed that the bug types we discovered are treated differently with respect to the process developers follow to fix them.}
	The proposed root cause classification model reached an overall F-Measure, AUC-ROC, and MCC of 64\%, 74\%, and 72\%, respectively, showing good performance when adopted for the classification of the most diffused bug root causes.

	\revised{Our future research agenda focuses on improving the devised model and better characterizing bugs referring to different root causes. Furthermore, we plan to exploit the proposed classification in other contexts: for instance, we envision the proposed taxonomy to be successfully employed for post-mortem analysis of bugs, as argued by Thung \etal \cite{thung2012automatic}; at the same time, we will investigate whether bug prioritization approaches can benefit from information on the nature of bugs, \eg security issues might be considered more important than GUI-related ones.}

\balance
\bibliographystyle{acmsmall}
\bibliography{references}

\end{document}